\documentclass[%
aps,
pra,
superscriptaddress,
tightenlines,
showpacs,showkeys,
a4paper,
12pt,
notitlepage
]{revtex4-1}
\usepackage[english]{babel}
\usepackage{amssymb,amsmath,stmaryrd,array}

\usepackage{graphicx}
\usepackage{subfig}

\makeatletter

\newcommand{\sca}[2]{\ensuremath{\bigl({#1}\cdot{#2}\bigr)}}

\newcommand{\prt}[1]{\partial_{#1}}

\newcommand{\sign}{\mathop{\rm sign}\nolimits}
\renewcommand{\Re}{\mathop{\rm Re}\nolimits}

 \newcommand{\bs}[1]{\boldsymbol{#1}}
 \newcommand{\vc}[1]{\mathbf{#1}}
 
 \newcommand{\uvc}[1]{\hat{\mathbf{#1}}}
 \newcommand{\ubs}[1]{\hat{\boldsymbol{#1}}}
 \newcommand{\ind}[1]{\mathrm{#1}}

\newcommand{\dd}{\mathrm{d}}

\newcommand{\inc}{\mathrm{inc}}

\newcommand{\med}{\mathrm{m}}

\makeatother

\begin{document}
\DeclareGraphicsExtensions{.eps,.png,.pdf}
\title{
Mie scattering of Laguerre-Gaussian beams:
photonic nanojets and near-field optical vortices
}

\author{Alexei~D.~Kiselev}
\email[Email address: ]{kiselev@iop.kiev.ua}
\affiliation{%
 Institute of Physics of National Academy of Sciences of Ukraine,
 prospekt Nauki 46,
 03680 Ky\"{\i}v, Ukraine}

\author{Dmytro~O.~Plutenko}
\email[Email address: ]{dmplutenko@gmail.com}
\affiliation{%
 Institute of Physics of National Academy of Sciences of Ukraine,
 prospekt Nauki 46,
 03680 Ky\"{\i}v, Ukraine}

%  \author{Vladimir~G.~Chigrinov}
%  \email[Email address: ]{eechigr@ust.hk}
% \affiliation{%
%  Hong Kong University of Science and Technology,
%  Clear Water Bay, Kowloon, Hong Kong
%  }

\date{\today}

\begin{abstract}
We study Mie light scattering of Laguerre-Gaussian (LG) beams
remodelled using the method of far-field matching.
The theoretical results are applied to examine
the optical field in the near-field region
for purely azimuthal LG beams characterized by
the nonzero azimuthal mode number $m_{\ind{LG}}$.
The mode number $m_{\ind{LG}}$ is found
to have a profound effect on 
the morphology
of photonic nanojets and the near-field structure of optical vortices
associated with the components of the electric field. 
\end{abstract}


 \maketitle

%%%%%%%%%%%%%%
\section{Introduction}
\label{sec:intro}
%%%%%%%%%%%%%%

The problem of light scattering by particles of one medium embedded in
another has a long history, dating back more than a century to the
classical exact solution due to Mie~\cite{Mie:1908}. The Mie solution
applies to scattering by uniform spherical particles with isotropic
dielectric properties. 
The analysis of a Mie--type theory uses a systematic expansion of
the electromagnetic field over vector spherical 
harmonics~\cite{New,Bohr,Tsang:bk1:2000,Mishchenko:bk:2004,Doicu:bk:2006,Gouesbet:bk:2011}.  
The
specific form of the expansions is also known as 
the \textit{T}--matrix ansatz that 
has been widely used in the related problem of light scattering
by nonspherical particles~\cite{Mis:1996,Mish,Mishchenko:bk:2004}.
More recently this strategy has been
successfully applied to optically anisotropic particles
~\cite{Rot:1973,Arag:1990,Arag:1994,Kar:1997,Kis:2000:opt,Kis:pre:2002,
Geng:pre:2004,Novitsky:pra:2008,Qiu:lpr:2010}.

In its original form the Mie theory assumes that 
the scatterer is illuminated with 
a plane electromagnetic wave.
For laser beams, 
it is generally necessary 
to go beyond
the plane-wave approximation
that may severely break down
when the beam width 
becomes of order of 
the scatterer size.
The problem of
light scattering from arbitrary shaped
laser beams
has now a more than two decade 
long history~\cite{Grehan:aplopt:1986,Gouesbet:josaa:1988,Barton:jap:1988,Barton:jap:2:1989,Schaub:josaa:1992}
and has been the key subject of
the Mie--type theory~---~the so-called
\textit{generalized Lorenz--Mie theory} (GLMT)
~\cite{Lock:jqsrt:2009,Gouesbet:bk:2011}~---~extended 
to the case of arbitrary incident-beam scattering. 

Mathematically, in such generalization of the Mie theory,
the central and the most important task
is to describe illuminating beams in terms of
expansions over a set of basis wavefunctions
(for the spherical coordinate system, it is 
the multipole expansion
over the basis
of vector spherical wavefunctions).
In GLMT, a variety of formally exact
(the quadrature and double quadrature formulas)
and approximate (the finite series and localized approximations)
methods~\cite{Gouesbet:aplopt:1988} were developed to
evaluate the expansion coefficients that
are referred to as the
\textit{beam shape coefficients}
(for a recent review see Ref.~\cite{Gouesbet:jqsrt:2:2011}
and references therein).

The central problem with laser beams
is due to the fact that 
in their standard mathematical form
these beams are not radiation fields
which are solutions to Maxwell's equations.
Typically, the analytical treatment
of laser beams is performed using 
the paraxial approximation~\cite{Lax:pra:1975}
and the beams are described as 
pseudo-fields
which are only approximate solutions
of the vector Helmholtz equation
(higher order corrections can be used to improve 
the accuracy of 
the paraxial approximation~\cite{Lax:pra:1975,Davis:pra:1979}).

Unfortunately, multipole expansions do not exist
for such approximate pseudo-fields.
Therefore, some remodelling procedure must be invoked
to obtain a real radiation field
which can be regarded as an approximation
to the original paraxial beam.

The basic concept that might be called \textit{matching the fields on 
a surface} lies at the heart of various traditional approaches to 
the laser beam remodelling
and is based on the assumption 
that there is a surface where 
the actual incident field is equal to the paraxial field.
Examples of physically reasonable and natural
choice are
scatterer-independent matching surfaces
such as
a far-field sphere~\cite{Nieminen:jqsrt:2003}, 
the focal plane 
(for beams with well-defined focal planes)~\cite{Nieminen:jqsrt:2003,Bareil:josaa:2013},
and a Gaussian reference sphere representing 
a lens~\cite{Hoang:josaa:2012}.
 Given the paraxial field distribution
on the matching surface,
the beam shape coefficients
can be evaluated using
either numerical integration
or the one-point matching method~\cite{Nieminen:jqsrt:2003}.

An alternative approach is to
describe analytically propagation of a laser beam,
which is known in the paraxial limit,
without recourse to the paraxial approximation.
In Refs.~\cite{Barnett:optcomm:1994,Duan:josaa:2005,Ness:jmo:2006,Zhou:ol:2006,Zhou:olt:2008}
this strategy has been applied
to the important case of Laguerre--Gaussian (LG) beams
using different methods such as
the vectorial Rayleigh--Sommerfeld
formulas~\cite{Duan:josaa:2005,Zhou:olt:2008},
the vector angular spectrum method~\cite{Zhou:ol:2006},
approximating LG beams by nonparaxial beams with (near) cylindrical
symmetry~\cite{Barnett:optcomm:1994,Ness:jmo:2006}.

The nonparaxial beams are solutions of Maxwell's equations
and the beam shape coefficients can be computed
using the methods of GLMT.
In recent studies of light scattering by spherical and spheroidal particles
illuminated with LG beams~\cite{Torok:optexp:2007,Jiang:jopt:2012}, 
the analytical results of Ref.~\cite{Ness:jmo:2006}
were used to calculate the beam shape coefficients.

It is now well known~\cite{Allen:bk:2003} that 
LG beams represent
optical vortex beams 
that carry angular momentum of two kinds:
spin angular momentum
associated with the polarization state of the beam
and orbital angular momentum
related to spatial variations of the field.
These variations derive from the helical structure of
the wavefronts comprising the beam or, equivalently,
from a phase singularity at the beam axis.
The topological charge characterizing the phase singularity
and associated orbital angular momentum
gives rise to distinctive phenomena
such as soliton generation~\cite{Kivshar:review:2005},
entanglement of photon quantum states,
orbital angular momentum
exchange with atoms and molecules
(in addition to the collection of papers~\cite{Allen:bk:2003},
see reviews in Ref.~\cite{Andrews:bk:2008}),
rotation and orbital motion of spherical particles 
illuminated with LG beams~\cite{Simpson:josaa:2009,Simpson:josaa:2010}.

In this paper the problem of light scattering 
from LG beams that represent 
laser beams exhibiting a helical phase front
and carrying a phase singularity will be of our primary interest.
In our calculations we shall 
follow Ref.~\cite{Kis:pre:2002}
and use the $T$--matrix approach
in which the far-field matching method
is combined with  the results for nonparaxial propagation of
LG beams~\cite{Zhou:ol:2006,Zhou:olt:2008}.
Our goal is to examine the near-field structure
of electromagnetic field depending on the parameters
characterizing both the beam and the scatterer. 

This structure has recently attracted
considerable attention
that was stimulated 
by an upsurge of interest to 
the so-called \textit{photonic nanojets}
and their applications (for a review see Ref.~\cite{Heifetz:jctn:2009}).
These nanojets were originally 
identified in finite-difference-time-domain 
simulations~\cite{Chen:optexp:2004,Li:optexp:2005}  
as narrow, high-intensity
electromagnetic beams 
that
propagate into background medium
from the shadow-side surface
of a plane-wave illuminated
dielectric microcylinder~\cite{Chen:optexp:2004} 
or microsphere~\cite{Li:optexp:2005}
of diameter greater than the illuminating
wavelength.
In other words, a photonic nanojet can be regarded as
a localized, subdiffractional, non-evanescent
light focus propagating along the line of incidence.

The bulk of 
theoretical 
studies devoted
to
nanojets~\cite{Lecler:ol:2005,Devilez:optexp:2008,Geints:optcom:2010,Ding:col:2010,
  Geints:josab:2012,Guo:optexp:2013} 
has been predominantly focused on the case of plane-wave 
illumination.
In this paper we intend to fill the gap.

The layout of the paper is as follows.  
In Sec.~\ref{sec:t-matrix-approach},
we describe our theoretical approach
and then, in Sec.~\ref{sec:incident-waves}, 
we obtain the analytical results
for the beam shape coefficients
of LG beams.
The numerical procedure and
the results of numerical computations
representing the near-field intensity distributions
and phase maps of electric field components
for purely azimuthal LG beams
are presented in Sec.~\ref{sec:results}.

Finally, in Sec.~\ref{sec:conclusions},
we present our results and make some concluding
remarks.

% Experiments on photonic
% nanojets~\cite{Heifetz:apl:2006,Ferrand:optexp:2008}

% sensitivity of Mie scattering to topological
% charge~\cite{Garbin:njp:2009}

% and 
% the effect of angular-momentum-induced transparency
% in Mie scattering of a circularly polarized, purely azimuthal LG beam~\cite{Rury:pra:2012}.

% Anomalous light scattering
% and giant absorption of light were discussed in 
% Refs.~\cite{Tribelsky:plr:2006,Tribelsky:epl:2011,Tribelsky:njp:2012,Fan:optexp:2010,Qiu:lpr:2010}.

%%%%%%%%%%%%%%%%%%%%%
\section{\textit{T}--matrix formulation of Lorenz--Mie theory}
\label{sec:t-matrix-approach}
%%%%%%%%%%%%%%%%%%%%%

We consider scattering by a spherical particle of radius $R_p$
embedded in a uniform isotropic dielectric medium with dielectric
constant $\epsilon_{\ind{med}}$ and magnetic
permeability $\mu_{\ind{med}}$.
The dielectric constant and magnetic
permittivity of the particle are $\epsilon_p$ and $\mu_p$,
respectively.

In this subsection we remind the reader about the relationship between
Maxwell's equations in the region of a scatterer and the formulation
of scattering properties in terms of the
\textit{T}--matrix~\cite{New,Mishchenko:bk:2004}. 
Our formulation is slightly non-standard
and closely follows to the line of our
presentation given in Ref.~\cite{Kis:pre:2002}. 
% Some technical details, which can be omitted at first
% reading, have been relegated to the appendices.

We shall need to write the  Maxwell equations 
for a harmonic electromagnetic wave
(time--dependent factor is $\exp\{-i\omega t\}$) in the
form:
\begin{subequations}
  \label{eq:maxwell}
\begin{align}
-i k_i^{-1}\,
\bs{\nabla}\times\vc{E}&=\frac{\mu_i}{n_i} \vc{H}\, ,
\label{eq:maxwell1}\\
i k_i^{-1}\,
\bs{\nabla}\times\vc{H}&= \frac{n_i}{\mu_i} \vc{E},
\quad
i=
\begin{cases}
  \ind{med}, & r>R_p\\
p, & r<R_p
\end{cases}
\label{eq:maxwell2}
\end{align}
\end{subequations}
where 
$n_{\ind{med}}=\sqrt{\epsilon_{\ind{med}}\mu_{\ind{med}}}$
is the refractive index outside the scatterer
(in the ambient medium),
where $r>R_p$ ($i=\ind{med}$)
and 
$k_i=k_{\ind{med}}=n_{\ind{med}}k_{\ind{vac}}$
($k_{\ind{vac}}=\omega/c=2\pi/\lambda$ is the free--space wavenumber);
$n_{p}=\sqrt{\epsilon_{p}\mu_{p}}$ 
is the refractive index
for the region inside the spherical particle (scatterer), 
where $r<R_p$ ($i=p$) and $k_i=k_p=n_pk_{\ind{vac}}$.

%%%%%%%%%%%%%%%%%%%%
\subsection{Vector spherical harmonics and Wigner \textit{D} functions}
\label{subsec:vsh-wigner-d-funcs}
%%%%%%%%%%%%%%%%%%%
  
The electromagnetic field can 
always be expanded
using the vector spherical harmonic basis,  
$\vc{Y}_{j+\delta\, j\, m}(\phi,\theta)\equiv
\vc{Y}_{j+\delta\, j\, m}(\uvc{r})$ 
($\delta=0,\pm 1$)~\cite{Biedenharn:bk:1981},
as follows:
\begin{subequations}
\label{eq:spher}
\begin{align}
& \vc{E}=\sum_{jm}\vc{E}_{jm}=\sum_{jm}\left[
       p^{(0)}_{jm}(r) \vc{Y}^{(0)}_{jm}(\uvc{r})+
       p^{(e)}_{jm}(r) \vc{Y}^{(e)}_{jm}(\uvc{r})+
       p^{(m)}_{jm}(r) \vc{Y}^{(m)}_{jm}(\uvc{r})\right]\:, 
\label{eq:e_spher}\\
& \vc{H}=\sum_{jm}\vc{H}_{jm}=\sum_{jm}\left[
       q^{(0)}_{jm}(r) \vc{Y}^{(0)}_{jm}(\uvc{r})+
       q^{(e)}_{jm}(r) \vc{Y}^{(e)}_{jm}(\uvc{r})+
       q^{(m)}_{jm}(r) \vc{Y}^{(m)}_{jm}(\uvc{r})\right]\:,
\label{eq:h_spher}
\end{align}
\end{subequations}
where
$\vc{Y}^{(m)}_{jm}=\vc{Y}_{j\,j\,m}$ and 
$\vc{Y}^{(e)}_{jm}=[j/(2j+1)]^{1/2}\vc{Y}_{j+1\,j\,m}+
[(j+1)/(2j+1)]^{1/2}\vc{Y}_{j-1\,j\,m}$ are electric and 
magnetic harmonics respectively, and 
$\vc{Y}^{(0)}_{jm}=[j/(2j+1)]^{1/2}\vc{Y}_{j-1\,j\,m}-
[(j+1)/(2j+1)]^{1/2}\vc{Y}_{j+1\,j\,m}$ are  longitudinal harmonics.
In Ref.~\cite{Kis:pre:2002},
it was shown that the spherical harmonics
can be conveniently expressed in terms of the Wigner 
\textit{D}--functions~\cite{Biedenharn:bk:1981,Varshalovich:bk:1988} 
as follows
\begin{subequations}
\label{eq:Y_D}
\begin{align}
& \vc{Y}_{jm}^{(m)}(\uvc{r})=
N_j /\sqrt{2}\left\{
D_{m,\,-1}^{j\,*}(\uvc{r})\,\vc{e}_{-1}(\uvc{r})-
D_{m,\,1}^{j\,*}(\uvc{r})\,\vc{e}_{+1}(\uvc{r})
\right\}\, ,
  \label{eq:Ym_D}\\
& \vc{Y}_{jm}^{(e)}(\uvc{r})=
N_j /\sqrt{2} \left\{
D_{m,\,-1}^{j\,*}(\uvc{r})\,\vc{e}_{-1}(\uvc{r})+
D_{m,\,1}^{j\,*}(\uvc{r})\,\vc{e}_{+1}(\uvc{r})
\right\}\, ,
  \label{eq:Ye_D}\\
& \vc{Y}_{jm}^{(0)}(\uvc{r})=
N_j
D_{m,\, 0}^{j\,*}(\uvc{r})\,\vc{e}_{0}(\uvc{r})
= Y_{jm}(\uvc{r})\uvc{r},
\quad
N_j=[(2j+1)/4\pi]^{1/2},
  \label{eq:Y0_D}
\end{align}
\end{subequations}
where 
$\vc{e}_{\pm 1}(\uvc{r})=
\mp (\vc{e}_x(\uvc{r})\pm i \vc{e}_y(\uvc{r}))/\sqrt{2}$;
$\vc{e}_x(\uvc{r})\equiv \ubs{\vartheta} =
(\cos\theta\cos\phi, \cos\theta\sin\phi, -\sin\theta)$,
$\vc{e}_y(\uvc{r})\equiv\ubs{\varphi}=(-\sin\phi, \cos\phi, 0)$
are the unit vectors tangential to the sphere;
$\phi$ ($\theta$) is the azimuthal (polar) angle of the unit vector 
$\uvc{r}=\vc{r}/r=
(\sin\theta\cos\phi, \sin\theta\sin\phi, \cos\theta)
\equiv\vc{e}_0(\uvc{r})\equiv\vc{e}_z(\uvc{r})$.
(Hats will denote unit vectors and an asterisk will indicate 
complex conjugation.)

Note that,
for the irreducible representation
of the rotation group with the angular number
$j$, 
the \textit{D}-functions,
$D_{m\nu}^{\,j}(\alpha,\beta,\gamma)=\exp(-i m \alpha)
d_{m\mu}^{\,j}(\beta) \exp(-i \mu \gamma)$, 
give the elements of the rotation matrix
parametrized by the three Euler 
angles~\cite{Biedenharn:bk:1981,Varshalovich:bk:1988}:
$\alpha$, $\beta$ and $\gamma$.
In formulas~\eqref{eq:Y_D} and throughout this paper,
we assume that
$\gamma=0$ and 
$D_{m\nu}^{\,j}(\uvc{r})\equiv D_{m\nu}^{\,j}(\phi,\theta,0)$.
These \textit{D}-functions
meet the following
orthogonality relations~\cite{Biedenharn:bk:1981,Varshalovich:bk:1988} 
\begin{equation}
  \label{eq:D-orth}
  \langle D_{m\nu}^{\,j\,*}(\uvc{r})
D_{m'\nu}^{\,j'}(\uvc{r})\rangle_{\uvc{r}}=
\frac{4\pi}{2j+1}\,\delta_{jj'}\,\delta_{mm'}\, ,
\end{equation}
where
$\displaystyle
\langle\,f\,\rangle_{\uvc{r}}\equiv\int_0^{2\pi}\dd\phi
\int_0^{\pi}\sin\theta\dd\theta\,f(\uvc{r})$
and $f(\uvc{r})\equiv f(\phi,\theta)$. The orthogonality
condition~\eqref{eq:D-orth} and Eqs.~\eqref{eq:Y_D}
show that a set of vector spherical harmonics is 
orthonormal:
\begin{equation}
  \langle \vc{Y}_{jm}^{(\alpha)\,*}(\uvc{r})\cdot
\vc{Y}_{j'm'}^{(\beta)}(\uvc{r})
\rangle_{\uvc{r}}= \delta_{\alpha\beta}\,\delta_{jj'}\,\delta_{mm'}\, .
  \label{eq:Y-orth}
\end{equation}

We can now use the relations~\cite{Biedenharn:bk:1981} 
\begin{align}
&
  \label{eq:Dm0_Yjm}
N_j D_{m\,0}^{\,j\,*}(\uvc{r})=\mathrm{Y}_{jm}(\uvc{r}),
  \\
&
  \label{eq:Dm1_Yjm}
N_j D_{m\,\pm 1}^{\,j\,*}(\uvc{r})=
n_j
\left[
\mp \prt{\theta} +\frac{i}{\sin\theta}\prt{\phi}
\right]
\mathrm{Y}_{jm}(\uvc{r}),
\quad
n_j\equiv [j(j+1)]^{-1/2}
\end{align}
where 
$\prt{x}$ stands for a derivative with respect to $x$ and
$\mathrm{Y}_{jm}(\uvc{r})$ is the normalized spherical function
\begin{align}
  \label{eq:Yjm-def}
  \mathrm{Y}_{jm}(\phi,\theta)=N_j  \exp(i m \phi) d^{\,j}_{m,
    0}(\theta)= (-1)^m
\sqrt{\frac{(2 j + 1) (j-m)!}{4 \pi (j+m)!}}\,\exp(i m \phi)\, P_j^{\,m}(\cos\theta)
\end{align}
expressed in terms of the associated Legendre polynomial
of degree $j$ and order $m$
\begin{align}
&
  \label{eq:Pjm}
  P_j^{\,m}(x)=
  \begin{cases}
    (-1)^m/(2^j j!)(1-x^2)^{m/2}\prt{x}^{j+m} (x^2-1)^j, & m>0\\
(-1)^{|m|} (j-|m|)!/(j+|m|)! P_j^{\,|m|}(x), & m<0
  \end{cases},
\end{align}
and derive the following expressions for
the magnetic and electric vector spherical functions
\begin{align}
&
\label{eq:Ym-in-Yjm}
\vc{Y}_{jm}^{(m)}(\uvc{r})=
-i n_j
\left[
[\prt{\theta}\ind{Y}_{jm}] \ubs{\varphi}
-i \frac{m}{\sin\theta} 
\ind{Y}_{jm}
\ubs{\vartheta}
\right]=
\notag
\\
&
n_j \vc{L} \ind{Y}_{jm}=-i \uvc{r}\times \vc{Y}_{jm}^{(e)},
\\
&
\label{eq:Ye-in-Yjm}
\vc{Y}_{jm}^{(e)}(\uvc{r})=
n_j
\left[
[\prt{\theta}\ind{Y}_{jm}]\ubs{\vartheta} 
+i \frac{m}{\sin\theta} 
\ind{Y}_{jm}
\ubs{\varphi}
\right]=
\notag
\\
&
n_j r \bs{\nabla} \ind{Y}_{jm} =-i \uvc{r}\times \vc{Y}_{jm}^{(m)},
\end{align}
where $\vc{L}$ is the operator of angular momentum
given by
\begin{align}
  \label{eq:ang-momentum}
  i \vc{L}=\vc{r}\times\bs{\nabla}
=\ubs{\varphi}\, \prt{\theta}
-\ubs{\vartheta}\, [\sin\theta]^{-1}\prt{\phi}.
\end{align}
Formulas~\eqref{eq:Ym-in-Yjm} and~\eqref{eq:Ye-in-Yjm} 
give the vector spherical harmonics~\eqref{eq:Y_D}
rewritten in the well-known standard form~\cite{Jacks:bk:1999}.

% (a number of relations for the vector spherical harmonics
% used throughout this paper
% are considered in Appendix~\ref{sec:vect-spher-harm}).

%%%%%%%%%%%%%%%%%%%%
\subsection{Wave functions and \textit{T}--matrix}
\label{subsec:t-matrix-ansatz}
%%%%%%%%%%%%%%%%%%%

The electric field~\eqref{eq:e_spher} is completely described by the coefficients
$\{p_{jm}^{(\lambda)}(r)\}$
and similarly the magnetic field~\eqref{eq:h_spher} is described by
$\{q_{jm}^{(\lambda)}(r)\}$ with $\lambda=\{0,e,m\}$.
In order to find the coefficient functions we  can use separation of
variables.  This implies that the expansions~\eqref{eq:spher} must be
inserted into Maxwell's equations~\eqref{eq:maxwell}. The coefficient
functions then can be derived by solving the resulting system of
equations.  In the simplest case of  an isotropic medium the coefficient
functions  
can be expressed in terms of spherical Bessel functions,
$j_j(x)=[\pi/(2x)]^{1/2} J_{j+1/2}(x)$, and spherical Hankel
functions~\cite{Abr}, $h_j^{(1,\,2)}(x)=[\pi/(2x)]^{1/2}
H_{j+1/2}^{(1,\,2)}(x)$, and their derivatives.

Alternatively, it is well-known 
(a discussion of the procedure can be found, e.g.,
in Ref.~\cite{Sarkar:pre:1997})
that solutions of the scalar
Helmholtz equation,
$(\bs{\nabla}^2+k^2)\psi(\vc{r})=0$,
taken in the form
\begin{align}
  \label{eq:psi_jm}
  \psi_{jm}^{(\alpha)}=n_j z_j^{(\alpha)}(\rho) \ind{Y}(\uvc{r}),
\quad n_j\equiv [j(j+1)]^{-1/2},
\end{align}
where $\rho=k r$ and 
$z_j^{(\alpha)}(\rho)$ is either a spherical Bessel or Hankel function,
can be used to obtain the following solenoidal solutions of the
vector Helmholtz equation,
$\bs{\nabla}\times[\bs{\nabla}\times\bs{\Psi}]=k^2\bs{\Psi}$: 
\begin{align}
&
  \label{eq:M_jm}
  \vc{M}_{jm}^{(\alpha)}(\rho,\uvc{r})=\vc{L}\psi_{jm}^{(\alpha)}=z_j^{(\alpha)}(\rho)\vc{Y}_{jm}^{(m)}(\uvc{r}),
\\
&
  \label{eq:N_jm}
\vc{N}_{jm}^{(\alpha)}(\rho,\uvc{r})=-i k^{-1}\,\bs{\nabla}\times\vc{M}_{jm}^{(\alpha)}=
\frac{\sqrt{j(j+1)}}{\rho}\,
z_j^{(\alpha)}(\rho)\, \vc{Y}_{jm}^{(0)}(\uvc{r})+
D z_j^{(\alpha)}(\rho) \vc{Y}_{jm}^{(e)}(\uvc{r}),
\end{align}
where $Df(x)\equiv x^{-1}\prt{x}(xf(x))$.
The vector wave functions, $\vc{M}_{jm}^{(\alpha)}$ and
$\vc{N}_{jm}^{(\alpha)}$,
are linked through the identities 
\begin{align}
  \label{eq:curl_NM}
  -i \bs{\nabla}\times\vc{M}_{jm}^{(\alpha)}= k \vc{N}_{jm}^{(\alpha)},\quad
  i \bs{\nabla}\times\vc{N}_{jm}^{(\alpha)}= k \vc{M}_{jm}^{(\alpha)}
\end{align}
and their linear combination represents the expansions~\eqref{eq:spher}
over the vector spherical harmonics.

There are three cases of these expansions that are of particular
interest. They correspond to the incident wave,
$\{\vc{E}_{\ind{inc}},\vc{H}_{\ind{inc}}\}$, the outgoing scattered wave,
$\{\vc{E}_{\ind{sca}},\vc{H}_{\ind{sca}}\}$
and the electromagnetic field inside the scatterer,
$\{\vc{E}_{p},\vc{H}_{p}\}$:
\begin{subequations}
\label{eq:EH}
\begin{align}
& 
\vc{E}_{\alpha}=
\sum_{jm}
\bigl[
\alpha_{jm}^{(\alpha)}\vc{M}_{jm}^{(\alpha)}(\rho_{i},\uvc{r})+
\beta_{jm}^{(\alpha)}\vc{N}_{jm}^{(\alpha)}(\rho_{i},\uvc{r})
\bigr],
\quad
\alpha\in\{\ind{inc}, \ind{sca}, p\}
\label{eq:E_alpha}
\\
& 
\vc{H}_{\alpha}=n_{i}/\mu_{i}\sum_{jm}
\bigl[
\alpha_{jm}^{(\alpha)}\vc{N}_{jm}^{(\alpha)}(\rho_{i},\uvc{r})-
\beta_{jm}^{(\alpha)}\vc{M}_{jm}^{(\alpha)}(\rho_{i},\uvc{r})
\bigr],
\label{eq:H_alpha}
\\
&
i=
\begin{cases}
  \ind{med}, & \alpha\in\{\ind{inc}, \ind{sca}\}\\
p, & \alpha=p
\end{cases},
\quad
z_j^{(\alpha)}(\rho_i)=
\begin{cases}
  j_j(\rho), & \alpha=\ind{inc}\\
h_j^{(1)}(\rho), & \alpha=\ind{sca}\\
j_j(\rho_p), & \alpha=p\\
\end{cases},
\end{align} 
\end{subequations}
where
$\rho_{\ind{med}}=k_{\ind{med}}r\equiv \rho$,
$\rho_{p}=k_{p} r\equiv n \rho$,
and $n=n_p/n_{\ind{med}}$ is 
the ratio of refractive indexes
also known as the optical contrast.

Thus outside the scatterer the electromagnetic field is a sum of the
incident wave field with $z_j^{(\ind{inc})}(\rho)=j_j(\rho)$
and the scattered waves with
$z_j^{(\ind{sca})}(\rho)=h_j^{(1)}(\rho)$ as required by the
Sommerfeld radiation condition.

In the far field region
($\rho\gg 1$), the asymptotic behaviour of the spherical Bessel
and Hankel functions is known~\cite{Abr}:
\begin{align}
&
  \label{eq:asymp-hankel1}
i^{j+1}h_j^{(1)}(\rho), i^{j}Dh_j^{(1)}(\rho)\sim\exp(i\rho)/\rho,
\\
&
  \label{eq:asymp-hankel2}
(-i)^{j+1}h_j^{(2)}(\rho), (-i)^{j}Dh_j^{(2)}(\rho)\sim\exp(- i\rho)/\rho,
\\
&
  \label{eq:asymp-bessel}
i^{j+1}j_j(\rho),i^{j+1}Dj_{j+1}(\rho)\sim
\bigl[\exp(i\rho)-(-1)^j \exp(-i\rho)
\bigr]/(2\rho).
\end{align}
So, the spherical Hankel functions of the first kind,
$h_j^{(1)}(\rho)$, describe the outgoing waves,
whereas those of the second kind, $h_j^{(2)}(\rho)$,
represent the incoming waves.
  
The incident field is the field that would exist
without a scatterer and therefore includes
both incoming and outgoing parts
(see Eq.~\eqref{eq:asymp-bessel})
because, when no scattering, 
what comes in must go outwards again.
As opposed to the spherical Hankel functions
that are singular at the origin, 
the incident wave field should be finite everywhere
and thus is described by the regular Bessel functions
$j_j(\rho)$.

Now the incident wave is characterized by amplitudes
$\alpha_{jm}^{(\ind{inc})}$, $\beta_{jm}^{(\ind{inc})}$ and the scattered
outgoing waves are similarly characterized by amplitudes
$\alpha_{jm}^{(\ind{sca})}$, $\beta_{jm}^{(\ind{sca})}$.  
So long as the scattering problem is
linear, the coefficients $\alpha_{jm}^{(\ind{sca})}$ and
$\beta_{jm}^{(\ind{sca})}$ can be written as linear combinations of
$\alpha_{jm}^{(\ind{inc})}$ and $\beta_{jm}^{(\ind{inc})}$:
\begin{widetext}
\begin{gather}
  \alpha_{jm}^{(\ind{sca})}=\sum_{j',m'}\left[\,
T_{jm,\,j'm'}^{\,11}\, \alpha_{j'm'}^{(\ind{inc})}+ 
T_{jm,\,j'm'}^{\,12}\,\beta_{j'm'}^{(\ind{inc})}
\,\right],
\notag
\\
\beta_{jm}^{(\ind{sca})}=\sum_{j',m'}\left[\,
T_{jm,\,j'm'}^{\,21}\, \alpha_{j'm'}^{(\ind{inc})}+ 
T_{jm,\,j'm'}^{\,22}\,\beta_{j'm'}^{(\ind{inc})}
\,\right]\, .
  \label{eq:matr}
\end{gather}
\end{widetext}
These formulae define the elements of the \textit{\textit{T}--matrix} in the most general case.

In general, the outgoing wave with angular momentum index $j$ arises
from ingoing waves of all other indices $j'$. In such cases we say
that the scattering process mixes angular momenta~\cite{Mis:1996}.
The light scattering from uniformly anisotropic 
scatterers~\cite{Kis:pre:2002,Kis:mclc:2002}
provides an example of such a scattering process.
In simpler scattering processes, by contrast, such angular momentum
mixing does not take place. Many quantum scattering processes and
classical Mie scattering belong to this category.  
For example, radial anisotropy keeps intact spherical symmetry of 
the scatterer~\cite{Rot:1973,Kis:pre:2002,Qiu:lpr:2010}.
The \textit{T}--matrix of a spherically symmetric scatterer is diagonal
over the angular momenta and the azimuthal numbers:
$T_{jj',mm'}^{nn'}=\delta_{jj'}\delta_{mm'} T_{j}^{nn'}$.

In order to calculate the elements of \textit{T}-matrix
and the coefficients $\alpha_{jm}^{(p)}$ and $\beta_{jm}^{(p)}$, we need to use
continuity of the tangential components of the electric and magnetic
fields as boundary conditions at $r=R_p$ 
($\rho=k_{\ind{med}} R_p\equiv x$).

So, the coefficients of the expansion for the wave field inside the
scatterer, $\alpha_{jm}^{(p)}$ and $\alpha_{jm}^{(p)}$, 
are expressed in terms of the coefficients
describing the incident light as follows

\begin{align}
&
  \label{eq:mie-alp-p}
 i \alpha_{jm}^{(p)}=
\frac{\alpha_{jm}^{(\ind{inc})}}{%
\mu^{-1} v_j(x) u_j^{\prime}(n x)
-n^{-1}v_j^{\prime}(x) u_j(n x)
},
\quad \mu=\mu_p/\mu_{\ind{med}},
\\
&
  \label{eq:mie-bet-p}
i \beta_{jm}^{(p)}=
\frac{\beta_{jm}^{(\ind{inc})}}{%
n^{-1}v_j(x) u_j^{\prime}(n x)-
\mu^{-1} v_j^{\prime}(x) u_j(n x)
},
\quad
n=n_p/n_{\ind{med}},  
\end{align}
where
$x=k_{\ind{med}} R_p$,
$u_j(x)= x j_j(x)$
and $v_j(x)= x h_j^{(1)}(x)$.
The similar result relating the scattered wave and the incident wave
\begin{align}
&
  \label{eq:mie-alp-sca}
  \alpha_{jm}^{(\ind{sca})}=
T_j^{11}\alpha_{jm}^{(\ind{inc})}=
\frac{n^{-1}u_j^{\prime}(x) u_j(n x)-\mu^{-1} u_j(x)
  u_j^{\prime}(n x)}{%
\mu^{-1} v_j(x) u_j^{\prime}(n x)
-n^{-1}v_j^{\prime}(x) u_j(n x)
}\alpha_{jm}^{(\ind{inc})},
\\
&
  \label{eq:mie-bet-sca}
  \beta_{jm}^{(\ind{sca})}=
T_j^{22}\beta_{jm}^{(\ind{inc})}=
\frac{
\mu^{-1} u_j(x) u_j^{\prime}(n x)-
n^{-1}u_j^{\prime}(x) u_j(n x)}{%
n^{-1}v_j(x) u_j^{\prime}(n x)-
\mu^{-1} v_j^{\prime}(x) u_j(n x)
}\beta_{jm}^{(\ind{inc})},
\end{align}
defines the \textit{T}-matrix for the simplest case of a spherically symmetric scatterer.
In addition, 
since the parity of electric and magnetic harmonics  
with respect to the spatial inversion $\uvc{r}\to -\uvc{r}$
($\{\phi,\theta\}\to \{\phi+\pi,\pi-\theta\}$)
is different
\begin{align}
  \label{eq:parity}
  \vc{Y}_{jm}^{(m)}(-\uvc{r})=(-1)^j \vc{Y}_{jm}^{(m)}(\uvc{r}),\quad
   \vc{Y}_{jm}^{(e)}(-\uvc{r})=(-1)^{j+1} \vc{Y}_{jm}^{(e)}(\uvc{r}),
\end{align}
where $f(\uvc{r})\equiv f(\phi,\theta)$ and $f(-\uvc{r})\equiv f(\phi+\pi,\pi-\theta)$,
they do not mix provided the mirror symmetry has not been broken. 
In this case the \textit{T}-matrix is diagonal and
$T_j^{12}=T_j^{21}=0$. The diagonal elements 
$T_j^{11}$ and $T_j^{22}$ are also called the \textit{Mie coefficients}.

%%%%%%%%%%%%%%%%%%%%
\section{Incident wave beams}
\label{sec:incident-waves}
%%%%%%%%%%%%%%%%%%%%

The formulas~\eqref{eq:mie-alp-p}-~\eqref{eq:mie-bet-sca} are useful only if the expansion for the incident
light beam is known. 
First we briefly review the most studied and
fundamentally important case where the incident light
is represented by a plane wave.

%%%%%%%%%%%%%%%%%%%%
\subsection{Plane waves}
\label{subsec:plane-wave}
%%%%%%%%%%%%%%%%%%%%

The electric field of a transverse plane wave 
propagating along the direction specified by a unit vector
$\uvc{k}_{\ind{inc}}$ is
\begin{equation}
  \label{eq:polar_inc}
\vc{E}_{\ind{inc}}=\vc{E}^{(\ind{inc})}\exp(i\,\vc{k}_{\ind{inc}}\cdot\vc{r})\, ,
\quad
\vc{E}^{(\ind{inc})}=\sum_{\nu=\pm 1}
E_{\nu}^{(\ind{inc})}\vc{e}_{\nu}(\uvc{k}_{\ind{inc}})\, ,
\qquad
\vc{k}_{\ind{inc}}=k\uvc{k}_{\ind{inc}}\, .
\end{equation}
where
the basis vectors $\vc{e}_{\pm 1}(\uvc{k}_{\ind{inc}})$ are perpendicular to
$\uvc{k}_{\ind{inc}}$.
Then the vector version of 
the well known Rayleigh expansion (see, for example,~\cite{New})
\begin{equation}
  \label{eq:Rayleigh}
  \exp(i\,\rho \uvc{k}\cdot\uvc{r})=
4\pi\sum_{l=0}^{\infty}\sum_{m=-l}^{l} 
i^{\,l} j_l(\rho)\,Y_{lm}(\uvc{r})\,Y_{lm}^{*}(\uvc{k}),\quad
\rho\equiv kr
\end{equation}
which is given by
\begin{align}
  \label{eq:vector_Rayleigh}
  \vc{e}_{\nu}(\uvc{k})\exp[i\rho \sca{\uvc{k}}{\uvc{r}}]
=
\sum_{jm}
\alpha_j D_{m\nu}^{\,j}(\uvc{k})
\Bigl\{
i\nu \vc{M}_{jm}(\rho,\uvc{r})-\vc{N}_{jm}(\rho,\uvc{r})
\Bigr\},
\quad \nu=\pm 1,
\end{align}
where
$\alpha_j=i^{j+1}[2\pi(2j+1)]^{1/2}$,
immediately gives the expansion coefficients for the plane wave
\begin{align}
  \label{eq:coef_inc}
\alpha_{jm}^{(\ind{inc})}&= i
\alpha_j 
\sum_{\nu=\pm 1} 
D_{m \nu}^{j}(\uvc{k}_{\ind{inc}})\nu 
E_{\nu}^{(\ind{inc})},
\quad
\beta_{jm}^{(\ind{inc})} = -
\alpha_j 
\sum_{\nu=\pm 1} 
D_{m \nu}^{j}(\uvc{k}_{\ind{inc}})
E_{\nu}^{(\ind{inc})}\, , 
\end{align}
where $D_{mm'}^j$ is the Wigner $D$-function. 

In the far field region, the electric field of scattered wave is
related to the polarization vector of the plane wave through the
scattering amplitude matrix as follows~\cite{New,Ishim,Mis:1996}
\begin{align}
\label{eq:ampl_def}
E_{\nu}^{(\ind{sca})}\equiv
(\vc{e}_{\nu}^{\,*}(\uvc{k}_{\ind{sca}}),
\vc{E}_{\ind{sca}})=
\rho^{-1}\exp(i\rho)\sum_{\nu'=\pm 1} 
\vc{A}_{\nu\nu'}(\uvc{k}_{\ind{sca}},\uvc{k}_{\ind{inc}}) 
E_{\nu'}^{(\ind{inc})}\, ,\quad \nu=\pm 1\,
\end{align}
where
$\uvc{k}_{\ind{sca}}=\uvc{r}$.
For a spherically symmetric scatterer, the expression for the
scattering amplitude matrix in 
terms of \textit{T}-matrix is given by
\begin{subequations}
  \label{eq:ampl_spher_sym}
\begin{align}
\vc{A}_{\nu\nu'}(\uvc{k}_{\ind{sca}},\uvc{k}_{\ind{inc}})&=
\sum_j \vc{A}_{\nu\nu'}^{j}(\uvc{k}_{\ind{sca}},\uvc{k}_{\ind{inc}})=\notag\\
=&-i\sum_{j} (j+1/2)\tilde{D}_{\nu\nu'}^{j}
(\uvc{k}_{\ind{sca}},\uvc{k}_{\ind{inc}})\,
\left[\,
\nu\nu'\, T_j^{\,11}-i\nu\, T_j^{\,12}+i\nu'\, T_j^{\,21}+T_j^{\,22}\,
\right]\, ,
\label{eq:ampl_spher_a}\\
& \tilde{D}_{\nu\nu'}^{j}(\uvc{k}_{\ind{sca}},\uvc{k}_{\ind{inc}})=
\sum_m D_{m\nu}^{j\,*}(\uvc{k}_{\ind{sca}})
D_{m\nu'}^{j}(\uvc{k}_{\ind{inc}})\, .
\label{eq:ampl_spher_b}
\end{align}
\end{subequations}
Equation~\eqref{eq:ampl_spher_b} shows that
the scattering amplitude matrix~\eqref{eq:ampl_spher_a} 
depends only on the angle between $\uvc{k}_{\ind{inc}}$ and $\uvc{k}_{\ind{sca}}$.
All far-field scattering characteristics of the system can be computed 
from the scattering amplitude matrix.

%%%%%%%%%%%%%%%%%%%%
\subsection{Far-field matching}
\label{subsec:far-field}
%%%%%%%%%%%%%%%%%%%%

Now we consider a more general case where   
an incident electromagnetic wave is written as a  superposition of
propagating plane waves:
\begin{subequations}
\label{eq:EH_plane-w-comb}  
\begin{align}
&
\label{eq:E_plane-w-comb}
    \vc{E}_{\ind{inc}}(\vc{r})\equiv
  \vc{E}_{\ind{inc}}(\rho,\uvc{r})=\langle\exp(i\rho\, \uvc{k}\cdot\uvc{r})\,
  \vc{E}_{\ind{inc}}(\uvc{k})
\rangle_{\uvc{k}},
\quad
  \vc{E}_{\ind{inc}}(\uvc{k})=
\sum_{\nu=\pm 1}
E_{\nu}(\uvc{k})\,\vc{e}_{\nu}(\uvc{k}),
\\
&
\label{eq:H_plane-w-comb}
\vc{H}_{\ind{inc}}(\vc{r})\equiv  
  \vc{H}_{\ind{inc}}(\rho,\uvc{r})=\frac{n}{\mu}\,\langle\exp(i\rho\, \uvc{k}\cdot\uvc{r})\,
\bigl[
\uvc{k}\times
  \vc{E}_{\ind{inc}}(\uvc{k})
\bigr]\rangle_{\uvc{k}}\, ,
\end{align}
\end{subequations}
where 
$\displaystyle
\langle\,f\,\rangle_{\uvc{k}}\equiv\int_0^{2\pi}\dd\phi_k
\int_0^{\pi}\sin\theta_k\dd\theta_k\,f$.

Our first step is to examine asymptotic behavior
of the wave field~\eqref{eq:EH_plane-w-comb}
in the far-field region, $\rho\gg 1$.
The results can be easily obtained
by using the asymptotic formula
for a plane wave (see, e.g., \cite{Mishchenko:bk:2004})
\begin{align}
&
  \label{eq:exp-asympt}
 \exp(i\rho\, \uvc{k}\cdot\uvc{r})\sim
\frac{-2\pi i}{\rho}
\bigl[
\exp(i\rho)
\delta(\uvc{k}-\uvc{r})
-\exp(-i\rho)
\delta(\uvc{k}+\uvc{r})
\bigr]
\quad \text{ at } \rho\gg 1,
\end{align}
where 
$\delta(\uvc{k}\mp \uvc{r})$
is the solid angle Dirac $\delta$-function
symbolically defined through the expansion
\begin{align}
&
   \label{eq:delta-angular}
\delta(\uvc{k}\mp \uvc{r})=
\sum_{l=0}^{\infty}\sum_{m=-l}^{l} 
Y_{lm}(\pm \uvc{r})\,Y_{lm}^{*}(\uvc{k}).  
\end{align}
Applying the relation~\eqref{eq:exp-asympt}
to the plane wave superposition~\eqref{eq:E_plane-w-comb}
gives the electric field of the incident wave
in the far-field region
\begin{align}
&
  \label{eq:E_inc-asympt}
 \vc{E}_{\ind{inc}}(\rho,\uvc{r})\sim
 \vc{E}_{\ind{inc}}^{(\infty)}(\rho,\uvc{r})=
\frac{1}{\rho}
\bigl[
\exp(i\rho)
\vc{E}_{\ind{out}}(\uvc{r})
+\exp(-i\rho)
\vc{E}_{\ind{in}}(\uvc{r})
\bigr],
\\
&
  \label{eq:E_in-out}
\vc{E}_{\ind{in}}(\uvc{r})
=-\vc{E}_{\ind{out}}(-\uvc{r}),
\end{align}
where $\vc{E}_{\ind{out}}(\uvc{r})$ is the far-field angular
distribution for the outgoing part of the electric field
of the incident wave:
\begin{align}
  \label{eq:E_out}
\vc{E}_{\ind{out}}(\uvc{r})
=-2\pi i\,
\vc{E}_{\ind{inc}}(\uvc{r})=
 E_{\theta}^{(\ind{out})}(\uvc{r})\,
\vc{e}_{\theta}(\uvc{r})
+
E_{\phi}^{(\ind{out})}(\uvc{r})\, 
\vc{e}_{\phi}(\uvc{r}),
\end{align}
whereas the incoming part of the incident wave
is described by the far-field angular distribution
$\vc{E}_{\ind{in}}(\uvc{r})$.

The result for the far-field distribution of 
the magnetic field~\eqref{eq:H_plane-w-comb}
can be written in the similar form:
\begin{align}
&
  \label{eq:H_inc-asympt}
 \vc{H}_{\ind{inc}}(\rho,\uvc{r})\sim
\vc{H}_{\ind{inc}}^{(\infty)}(\rho,\uvc{r})=
\frac{1}{\rho}
\bigl[
\exp(i\rho)
\vc{H}_{\ind{out}}(\uvc{r})
+\exp(-i\rho)
\vc{H}_{\ind{in}}(\uvc{r})
\bigr],
\\
&
  \label{eq:H_in-out}
\vc{H}_{\ind{in}}(\uvc{r})
=-\vc{H}_{\ind{out}}(-\uvc{r}),
\\
&
  \label{eq:H_in-H_out}
\mu/n\,\vc{H}_{\ind{out}}(\uvc{r})=
\uvc{r}\times \vc{E}_{\ind{out}}(\uvc{r}),
\quad
\mu/n\,\vc{H}_{\ind{in}}(\uvc{r})=
\uvc{r}\times \vc{E}_{\ind{out}}(-\uvc{r}).
\end{align}
Formulas~\eqref{eq:E_inc-asympt}-\eqref{eq:H_in-H_out}
explicitly show that, in the far-field region,
the incident wave field is defined by
the angular distribution of the outgoing wave~\eqref{eq:E_out}.
In particular, from these formulas,
it is not difficult to obtain the far-field expression for 
the Poynting vector of the incident wave
$\vc{S}_{\ind{inc}}=c/(8\pi)\Re(\vc{E}_{\ind{inc}}\times\vc{H}_{\ind{inc}}^{\,*})$
\begin{align}
&
  \label{eq:S_inc-asympt}
 \vc{S}_{\ind{inc}}(\rho,\uvc{r})\sim
\vc{S}_{\ind{inc}}^{(\infty)}(\rho,\uvc{r})=
\rho^{-2}
\bigl\{
\vc{S}_{\ind{in}}(\uvc{r})
+\vc{S}_{\ind{out}}(\uvc{r})
\bigr\},
\\
&
  \label{eq:S_in-out}
\vc{S}_{\ind{in}}(\uvc{r})
=-\vc{S}_{\ind{out}}(-\uvc{r}),
\quad
\mu/n\,\vc{S}_{\ind{out}}(\uvc{r})=
c/(8\pi)\,
|\vc{E}_{\ind{out}}(\uvc{r})|^2\,\uvc{r} ,
\end{align}
where
$|\vc{E}_{\ind{out}}(\uvc{r})|^2=(\vc{E}_{\ind{out}}(\uvc{r})\cdot\vc{E}_{\ind{out}}^{\,*}(\uvc{r}))$.
From this expression it immediately follows that 
the flux of Poynting vector of the outgoing wave,
$\vc{S}_{\ind{out}}(\uvc{r})$,
through a sphere of sufficiently large radius
is exactly balanced by 
the flux of Poynting vector of the incoming wave,
$\vc{S}_{\ind{inc}}(\uvc{r})$.

Alternatively, the far-field distribution
of an incident light beam,
$\vc{E}_{\ind{out}}(\uvc{r})$,
can be found from the expansion
over the vector spherical harmonics~\eqref{eq:E_alpha}.  
The far-field asymptotics
for the vector wave functions
that enter the expansion for the
incident wave~\eqref{eq:EH}
\begin{align}
&
  \label{eq:M_inc-asympt}
 \vc{M}_{jm}^{(\ind{inc})}(\rho,\uvc{r})\sim
\frac{(-i)^{j+1}}{2 \rho}
\bigl[
\exp(i\rho)
\vc{Y}_{jm}^{(m)}(\uvc{r})
-\exp(-i\rho)
\vc{Y}_{jm}^{(m)}(-\uvc{r})
\bigr],
\\
&
  \label{eq:N_inc-asympt}
 \vc{N}_{jm}^{(\ind{inc})}(\rho,\uvc{r})\sim
\frac{(-i)^j}{2 \rho}
\bigl[
\exp(i\rho)
\vc{Y}_{jm}^{(e)}(\uvc{r})
-\exp(-i\rho)
\vc{Y}_{jm}^{(e)}(-\uvc{r})
\bigr],
\end{align}
can be derived from Eqs.~\eqref{eq:M_jm}-\eqref{eq:N_jm}
with the help of the far-field relation~\eqref{eq:asymp-bessel}.
Substituting Eqs.~\eqref{eq:M_inc-asympt} and~\eqref{eq:N_inc-asympt}
into the expansion~\eqref{eq:E_alpha} gives the far-field distribution
of the form~\eqref{eq:E_inc-asympt}
with
\begin{align}
  \label{eq:E_out_expan}
 \vc{E}_{\ind{out}}(\uvc{r})=
2^{-1}
\sum_{jm} 
\Bigl[
(-i)^{j+1} \alpha_{jm}^{(\ind{inc})}
\vc{Y}_{jm}^{(m)}(\uvc{r})
+
(-i)^{j} \beta_{jm}^{(\ind{inc})}
\vc{Y}_{jm}^{(e)}(\uvc{r})
\Bigr].
\end{align}
The coefficients of the incident wave
can now be easily found as the Fourier coefficients 
of the far-field angular distribution, $\vc{E}_{\ind{out}}$,
expanded using the vector spherical harmonics basis~\eqref{eq:Y_D}.
The final result reads
\begin{subequations}
\label{eq:alp_beta_inc}
\begin{align}
&
  \label{eq:alp_inc}
\alpha_{jm}^{(\ind{inc})}= 2\, i^{j+1}
  \langle \vc{Y}_{jm}^{(m)\,*}(\uvc{r})\cdot
\vc{E}_{\ind{out}}(\uvc{r})
\rangle_{\uvc{r}}=
i\alpha_{j}\sum_{\nu=\pm 1}
\nu
\langle D_{m \nu}^{\,j}(\uvc{k})\,
E_{\nu}(\uvc{k})
\rangle_{\uvc{k}},
\\
&
  \label{eq:beta_inc}
 \beta_{jm}^{(\ind{inc})}= 2\, i^{j}
  \langle \vc{Y}_{jm}^{(e)\,*}(\uvc{r})\cdot
\vc{E}_{\ind{out}}(\uvc{r})
\rangle_{\uvc{r}}=
-\alpha_{j}\sum_{\nu=\pm 1}
\langle D_{m \nu}^{\,j}(\uvc{k})\,
E_{\nu}(\uvc{k})
\rangle_{\uvc{k}}.
\end{align}
\end{subequations}
A comparison between the expressions on the right hand side
of Eq.~\eqref{eq:alp_beta_inc}
and those for the plane wave~\eqref{eq:coef_inc}
shows that, in agreement with the representation~\eqref{eq:E_plane-w-comb}, 
the result for plane waves represents
the limiting case where the angular distribution is singular:
$E_{\nu}(\uvc{k})=E_{\nu}^{(\ind{inc})}\,\delta(\uvc{k}-\uvc{k}_{\ind{inc}})$.  

By using Eqs.~\eqref{eq:Ym-in-Yjm} and~\eqref{eq:Ye-in-Yjm}
formulas~\eqref{eq:alp_beta_inc} can be conveniently rewritten
in the explicit form
\begin{subequations}
\label{eq:alp_beta_inc2}
\begin{align}
&
  \label{eq:alp_inc2}
\alpha_{jm}^{(\ind{inc})}= 2 n_j\, i^{j+1}
  \langle \mathrm{Y}_{jm}^{\,*}(\uvc{r})\,
(\vc{L}
\cdot
\vc{E}_{\ind{out}}(\uvc{r}))
\rangle_{\uvc{r}}=
\notag
\\
&
2 n_j\, i^{j}
\int_{0}^{2\pi}\dd\phi\int_{0}^{\pi}\dd\theta\,
\mathrm{Y}_{jm}^{\,*}(\phi,\theta)
\Bigl[
\prt{\theta}(\sin\theta E_{\phi}^{(\ind{out})})
- \prt{\phi}E_{\theta}^{(\ind{out})}
\Bigr],
\\
&
  \label{eq:beta_inc2}
 \beta_{jm}^{(\ind{inc})}= -2 n_j\, i^{j}\,
  \langle \mathrm{Y}_{jm}^{\,*}(\uvc{r})\,
(r \bs{\nabla}
\cdot
\vc{E}_{\ind{out}}(\uvc{r}))
\rangle_{\uvc{r}}=
\notag
\\
&
-2 n_j\, i^{j}
\int_{0}^{2\pi}\dd\phi\int_{0}^{\pi}\dd\theta\,
\mathrm{Y}_{jm}^{\,*}(\phi,\theta)
\Bigl[
\prt{\theta}(\sin\theta E_{\theta}^{(\ind{out})})
+ \prt{\phi}E_{\phi}^{(\ind{out})}
\Bigr],
\end{align}
\end{subequations}
which might be useful for computational purposes.

We conclude this section with the remark concerning 
the effect of translation
\begin{align}
  \label{eq:shift_EH}
  \{\vc{E}_{\ind{inc}}(\vc{r}),\vc{H}_{\ind{inc}}(\vc{r})\}
\to
\{\vc{E}_{\ind{inc}}(\vc{r}-\vc{R}_d),\vc{H}_{\ind{inc}}(\vc{r}-\vc{R}_d)\}
\end{align}
on the far-field angular distribution~\eqref{eq:E_out}.
Note that,
under the action of transformation~\eqref{eq:shift_EH}, 
the focal plane is displaced 
from its initial position by the vector $\vc{R}_d$.
From Eqs.~\eqref{eq:EH_plane-w-comb}
and~\eqref{eq:E_out},
it follows that, for the far-field distribution~\eqref{eq:E_out},
translation results in the phase shift
\begin{align}
  \label{eq:shift_E_out}
  \vc{E}_{\ind{out}}(\uvc{r})
\to
\vc{E}_{\ind{out}}(\uvc{r})\exp(-i k R_r),
\end{align}
where $R_r=(\vc{R}_d\cdot\uvc{r})$
is the radial component of the displacement vector $\vc{R}_d$.

%%%%%%%%%%%%%%%%%%%%
\subsection{Laguerre--Gaussian beams}

\label{subsec:LG-beams}
%%%%%%%%%%%%%%%%%%%%

% Multipole expansion of laser
% beams~\cite{Devaney:jmath:1974,Sherman:jmath:1976,Sherman:josaa:1982,Clemmow:bk:1996,Doicu:optcomm:1997}

In the paraxial approximation,
the beams are described in terms of
scalar fields of the form: $u(\vc{r}) \exp(i k z)$,
where $u(\vc{r})$ is a solution
of the paraxial Helmholtz equation
\begin{align}
  \label{eq:parax-Helmholtz}
[\bs{\nabla}_{\perp}^2 + 2 i k \prt{z}] u = 0,
\quad
  \bs{\nabla}_{\perp}^2=\prt{x}^2+\prt{y}^2.
\end{align}
For LG beams, the solution can be conveniently
written in the cylindrical coordinate system,
$(r_{\perp},\phi,z)$, as follows
\begin{subequations}
  \label{eq:LG-scalar}
\begin{align}
&
  \label{eq:LG-u_mn}
  u_{nm}(r_{\perp},\phi,z)=
  |\sigma|^{-1}\psi_{nm}(\sqrt{2}r_{\perp}/w)
\exp\{
-r_{\perp}^2/(w_0^2\sigma)
+ i m \phi - i \gamma_{nm}
\},
\\
&
  \label{eq:LG-sigma-w}
\sigma\equiv\sigma(z)=1+i z/z_R,
\quad
w\equiv w(z)=w_0|\sigma|,
\\
&
  \label{eq:LG-gamma-psi}
\gamma_{nm}\equiv\gamma_{nm}(z)=(2n+m+1)\arctan(z/z_R),
\quad
\psi_{nm}(x)=x^{|m|} L_n^{|m|}(x^2),
\end{align}
\end{subequations}
where
$L_n^{m}$ is the generalized Laguerre polynomial given by~\cite{Grad:1980}
\begin{align}
  \label{eq:L_nm}
  L_n^m(x)=(n!)^{-1} x^{-m} \exp(x)\,  \prt{x}^{n}\,[x^{n+m}\exp(-x)],
\end{align}
$n$ ($m$) is the radial (azimuthal) mode number;
$w_0$ is the initial transverse Gaussian half-width
(the beam diameter at waist)
$z_R=k w_0^2/2=[2 k f^2]^{-1}$ is the Rayleigh range
and $f=[k w_0]^{-1}$.
Note that, in addition to the standard mathematical methods, 
the result~\eqref{eq:LG-scalar} can also be 
obtained using either the ladder operator technique~\cite{Nienhuis:pra:1993}
or the operator approach developed in Ref.~\cite{Enderlein:josa:2004}.

The problem studied in Refs.~\cite{Zhou:ol:2006,Duan:josaa:2005,Zhou:olt:2008}
deals with  the exact propagation of the optical
field in the half-space, $z>0$, when 
its transverse components at the initial (source) plane,
$z=0$, are known.
In Ref.~\cite{Zhou:ol:2006}, 
the results describing asymptotic behavior of the linearly polarized 
field
\begin{align}
  \label{eq:LG-init}
  \vc{E}(r_{\perp},\phi,0)=u_{nm}(r_{\perp},\phi,0)\,\uvc{x}=
  \psi_{nm}(\sqrt{2}r_{\perp}/w_0)
\exp\{
-r_{\perp}^2/w_0^2
+ i m \phi\}\,\uvc{x}
\end{align}
were derived using the angular spectrum representation
(Debye intergrals)
and comply with both
the results of rigorous mathematical analysis
performed in Ref.~\cite{Sherman:jmath:1976}
and those obtained using the vectorial Rayleigh-Sommerfeld
integrals~\cite{Duan:josaa:2005,Zhou:olt:2008}.
The resulting expression for the far-field angular distribution
can be written in the following form
\begin{subequations}
  \label{eq:E_ff_LG}
\begin{align}
&
  \label{eq:E_out_LG}
  \vc{E}_{\ind{out}}^{(\ind{LG})}(\phi,\theta)=E_{nm}(f^{-1}\sin\theta/\sqrt{2})\,
\exp(im \phi)
\vc{e}_{\ind{out}},
\\
&
 \label{eq:e_out}
\vc{e}_{\ind{out}}
=
\cos\phi\,\vc{e}_{\theta}(\uvc{r})-
\cos\theta\,\sin\phi\,\vc{e}_{\phi}(\uvc{r})
=
\cos\theta\,\uvc{x}-\sin\theta\,\cos\phi\,\uvc{z},
\\
&
 \label{eq:E_nm}
E_{nm}(x)=\frac{ x^m}{i^{2n+m+1} 2 f^{2}}\,
L_n^{m}(x^2)\,\exp(-x^2/2).
\end{align}
\end{subequations}

We can now combine the relations~\eqref{eq:E_out}
and~\eqref{eq:EH_plane-w-comb}
with the outgoing part of
the far-field distribution~\eqref{eq:E_out_LG}
to deduce the expression for the electric field
of the remodelled LG beam
\begin{align}
&
  \label{eq:E_inc_LG}
  \vc{E}_{\ind{inc}}^{(\ind{LG})}(\rho_\perp,\phi,\rho_z)=
  E_{x}^{(\ind{LG})}(\rho_\perp,\phi,\rho_z)\,\uvc{x}+
  E_{z}^{(\ind{LG})}(\rho_\perp,\phi,\rho_z)\,\uvc{z}=
\notag
\\
&
\frac{i}{2\pi}
\langle
\exp\left[
i(
\rho_\perp\sin\theta_k\cos(\phi-\phi_k)
+\rho_z\cos\theta_k
)
\right]
\,
\vc{E}_{\ind{out}}^{(\ind{LG})}(\uvc{k})
\rangle_{\uvc{k}},
\end{align}
where $\rho_\perp=k r_\perp$ and $\rho_z=k z$.
For computational purposes,
the electric field can be conveniently recast into 
the explicit form with the help of the identity~\cite{Abr}
\begin{align}
  \label{eq:rel-exp-cos}
  \exp[i x \cos\phi]
=J_0(x)+ 2\sum_{k=1}^{\infty} i^k J_k(x)\cos k\phi,
\end{align}
where $J_m(x)$ is the Bessel function
of the first kind of order $m$.
The final result reads
\begin{subequations}
  \label{eq:Exz_inc_LG}
\begin{align}
&
  \label{eq:Ex_inc_LG}
  E_x^{(\ind{LG})}=i^{m+1}\exp[i m \phi]
\int_{0}^{\pi/2}
J_m(\rho_\perp\sin\theta_k)
F_{nm}(\rho_z,\theta_k)
\cos\theta_k
\sin\theta_k
\dd\theta_k,
\\
&
  \label{eq:Ez_inc_LG}
  E_z^{(\ind{LG})}=i^{m}/2\sum_{\delta=\pm1}\delta\exp[i (m+\delta) \phi]
\int_{0}^{\pi/2}
J_{m+\delta}(\rho_\perp\sin\theta_k)
F_{nm}(\rho_z,\theta_k)
\sin^2\theta_k
\dd\theta_k,
\end{align}
\end{subequations}
where
$F_{nm}(\rho_z,\theta_k)\equiv \exp[i\rho_z\cos\theta_k]
E_{nm}(f^{-1}\sin\theta_k/\sqrt{2})$.

Note that, for the so-called cosine and sine LG beams,
similar expressions 
can be obtained from Eq.~\eqref{eq:Exz_inc_LG}
by replacing
the exponential factors, 
$\exp[i m \phi]$ in Eq.~\eqref{eq:Ex_inc_LG} and $\exp[i (m+\delta)
\phi]$ in Eq.~\eqref{eq:Ez_inc_LG},
with their real and imaginary parts, respectively. 
% For cosine and sine LG beams,
% the factor $\exp(im \phi)$
% that enters the left hand side of Eqs.~\eqref{eq:LG-init}
% and~\eqref{eq:E_out_LG}
% must be replaced with $\cos(m\phi)$ and $\sin(m\phi)$,
% respectively.
These beams are real-valued at the focal plane $z=0$
and might be called \textit{``dark LG beams''}
(in general, ``dark beams'' are nonuniform optical beams that contain either a
one-dimensional (1D) dark stripe or a two-dimensional (2D) dark hole
resulting from a phase singularity or an amplitude depression in their
optical field).

%%%%%%%%%%%%%%%%%%%%%%%
\section{Results and discussion}
\label{sec:results}
%%%%%%%%%%%%%%%%%%%%%%

In this section,
we 
present the results of numerical 
computations on the light scattering problem
for the case
where
the incident wave 
is represented by 
the remodelled LG beams~\eqref{eq:E_inc_LG}
with the vanishing radial mode number $n=0$ and
the non-negative azimuthal number
$m=m_{\ind{LG}}\ge 0$.
Such beams are also known as the 
\textit{purely azimuthal LG beams}~\cite{Rury:pra:2012}.

Substituting the far-field distribution~\eqref{eq:E_ff_LG}
into Eq.~\eqref{eq:alp_beta_inc2}
gives the beam shape coefficients of these beams
in the following form:
\begin{subequations}
\label{eq:alpha-beta-inc-rules}
\begin{align}
&
  \label{eq:alpha-inc-rules}
  \alpha_{jm}^{(\inc)}=
\alpha_{j,\,m_{\ind{LG}}}^{(+)}\,\delta_{m,\,m_{\ind{LG}}+1}+
\alpha_{j,\,m_{\ind{LG}}}^{(-)}\,\delta_{m,\,m_{\ind{LG}}-1},
\\
&
\label{eq:beta-inc-rules}
  \beta_{jm}^{(\inc)}=
\beta_{j,\,m_{\ind{LG}}}^{(+)}\,\delta_{m,\,m_{\ind{LG}}+1}+
\beta_{j,\,m_{\ind{LG}}}^{(-)}\,\delta_{m,\,m_{\ind{LG}}-1}.
\end{align}
\end{subequations}
Then 
the coefficients of expansions~\eqref{eq:EH}
describing scattered wave
and electromagnetic field inside the scatterer
can be evaluated from 
formulas~\eqref{eq:mie-alp-p}--~\eqref{eq:mie-bet-sca}.

\begin{figure*}[!tbh]
%\centering
%\resizebox{100mm}{!}{\includegraphics*{nanojet.eps}}
\includegraphics[width=120mm]{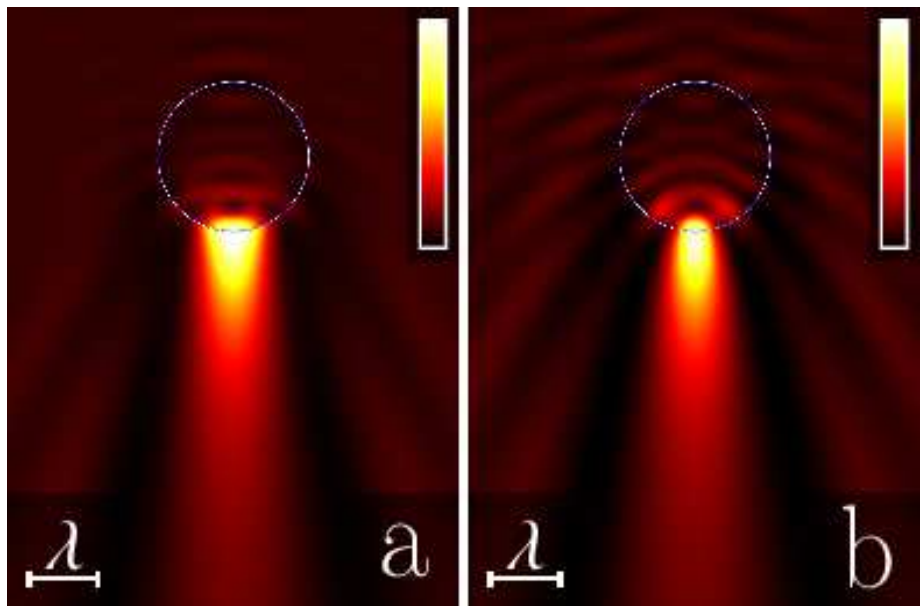}
\caption{%
Near-field intensity distributions 
of the total wavefield
in (a)~the $x-z$ plane
and (b)~the $y-z$ plane
for the LG beam with
$m_{\ind{LG}}=0$, 
$f=0.05$ and $|\vc{R}_d|=0$.
The parameters are:
$R_p=\lambda$ is the scatterer radius
and $n_p=1.3$ ($n_{\med}=1.0$)
is the refractive index inside (outside)
the particle. 
}
\label{fig:m0}
\end{figure*}

%%%%%%%%%%%%%%%%%%%%%%%
\subsection{Photonic nanojets}
\label{subsec:nonojets}
%%%%%%%%%%%%%%%%%%%%%%

For spherical particles
illuminated by plane waves,
formation of photonic nanojets and their structure
was previously discussed
in Refs.~\cite{Lecler:ol:2005,Devilez:optexp:2008,Geints:optcom:2010}.
Plane waves can be regarded as Gaussian beams
with $n=m_{\ind{LG}}=0$ and sufficiently small focusing 
parameter, $f\ll 1$, which is defined after Eq.~\eqref{eq:L_nm} 
through the ratio of wavelength, $\lambda$, 
and the beam diameter at waist, $w_0$,
$ f = (2\pi)^{-1} \lambda/w_0$.
This limiting case is illustrated in
Fig.~\ref{fig:m0} which shows the near-field intensity distributions
for the total light wavefield 
in both the $x-z$ and the $y-z$ planes
computed at $m_{\ind{LG}}=0$
and $f=0.05$ for the spherical particle of the radius $R_p=\lambda$ 
with the refractive index $n_p=1.3$ (water) located in the air
($n_{\ind{m}}=1$).

\begin{figure*}[!tbh]
%\centering
%\resizebox{100mm}{!}{\includegraphics*{nanojet.eps}}
\includegraphics[width=100mm]{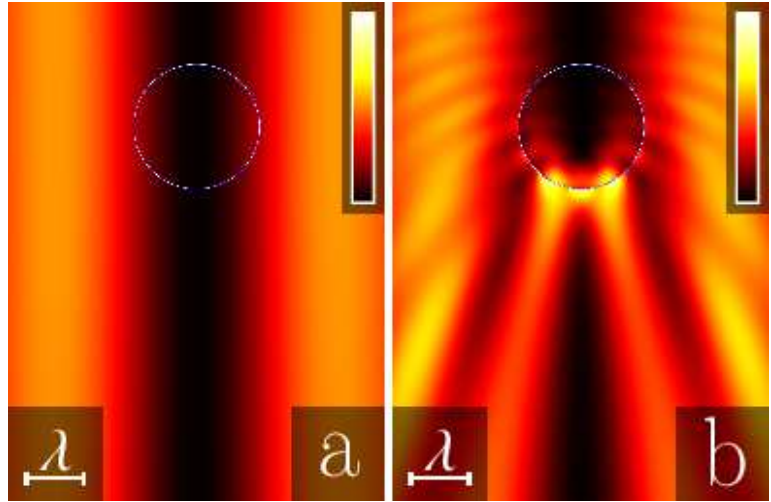}
\caption{%
Near-field intensity distribution in the $x-z$ plane
of  (a)~the incident wave beam
and (b)~the total wavefield
for the LG beam with
$m_{\ind{LG}}=1$, 
$f=0.05$ and $|\vc{R}_d|=0$.
Other parameters are
described in the caption of Fig.~\ref{fig:m0}.
}
\label{fig:m1}
\end{figure*}

It can be seen that the distributions
are characterized by the presence of elongated focusing zones
formed near the shadow surface of the scatterer.
The transverse size of these zones is smaller than the wavelength
of incident light, whereas their longitudinal size 
in the direction of incidence which is along the $z$ axis from top to
bottom is relatively large. 
Such a jetlike light structure is typical for the photonic nanojets.
The characteristic length and width of nanojets 
along with the peak intensity are known to
strongly depend on a number of factors such
as the scatterer size $R_p$, the particle absorption coefficient 
and the optical contrast ratio $n_p/n_{\ind{m}}$.
For microspheres,
the results of a comprehensive numerical analysis 
including the case of shell particles
are summarized in the recent paper~\cite{Geints:optcom:2010}.

Effects of
non-plane incident waves 
such as the laser beams 
on the structure of photonic nanojets
are much less studied.
Some theoretical results for tightly focused
Gaussian beams are reported
in Ref.~\cite{Devilez:optexp:2009} 
and the case of Bessel-Gauss beams
was studied experimentally in~\cite{Kim:optexp:2011}.

For the LG beams, we begin with the effects of the azimuthal
mode number and describe what happens to 
the near-field structure shown in Fig.~\ref{fig:m0}
when the azimuthal number takes the smallest non-zero value, 
$m_{\ind{LG}}=1$.
The latter represents the simplest case of 
an optical vortex beam in which, 
owing to the presence of phase singularity, 
the intensity of incident light at the beam axis
(the $z$ axis) vanishes (see Fig.~\ref{fig:m1}(a)).
From Fig.~\ref{fig:m1}, 
it can be seen that,
even though the bulk part of the scatterer
is in the low intensity region
surrounding the optical vortex,
the scattering process is efficient enough to produce
scattered waves 
that result in the formation of a pronounced 
jetlike photonic 
flux emerging from the particle shadow surface
(see Fig.~\ref{fig:m1}(b)).

\begin{figure*}[!htb]
%\centering
%\resizebox{100mm}{!}{\includegraphics*{nanojet.eps}}
\includegraphics[width=100mm]{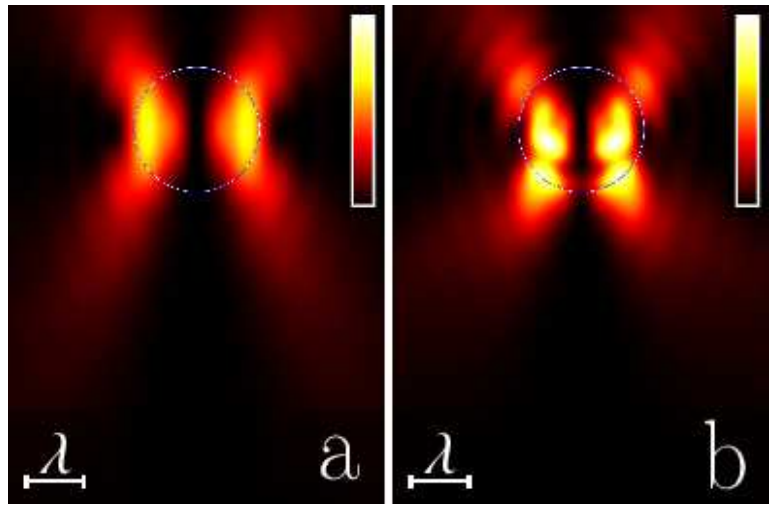}
\caption{%
Near-field intensity distribution in the $x-z$ plane
of  (a)~the incident wave beam
and (b)~the total wave field
for the LG beam with
$m_{\ind{LG}}=2$, $f=0.25$
and $|\vc{R}_d|=0$. 
}
\label{fig:m2_0}
\end{figure*}

A comparison between Fig.~\ref{fig:m1}(b) 
and Fig.~\ref{fig:m0}(a)
shows that
the three-peak structure of the photonic jet
formed at 
Mie scattering of the optical vortex LG beam
with $m_{\ind{LG}}=1$
significantly differs from
the well-known shape of the nanojet
at  $m_{\ind{LG}}=0$.
Interestingly,
similar to the case of Gaussian beams
with $m_{\ind{LG}}=0$,
the focusing zones at $m_{\ind{LG}}=1$
involve the beam axis
where one of the light intensity peaks is located.

\begin{figure*}[!htb]
%\centering
%\resizebox{100mm}{!}{\includegraphics*{nanojet.eps}}
\includegraphics[width=100mm]{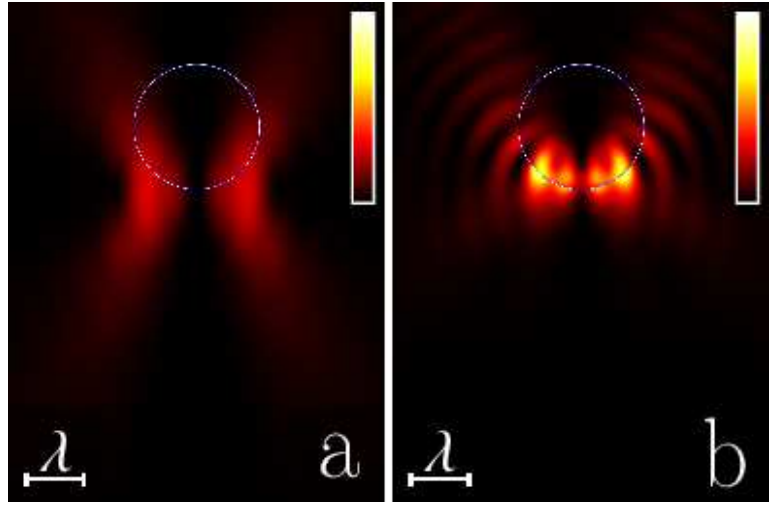}
\caption{%
Near-field intensity distribution in the $x-z$ plane
of  (a)~the incident wave beam
and (b)~the total wave field
for the LG beam with
$m_{\ind{LG}}=2$, $f=0.25$
and $\vc{R}_d=(0,0,\lambda)$. 
}
\label{fig:m2_1}
\end{figure*}

The results for tightly focused LG beams
with $m_{\ind{LG}}=2$ and $f=0.25$ 
are shown in Figs.~\ref{fig:m2_0}
and~\ref{fig:m2_1}.
When the displacement 
vector, $\vc{R}_d$ defined in 
Eqs.~\eqref{eq:shift_EH}
vanishes,
the focal (waist) plane of the incident LG beam
is $z=0$ and contains the center
of the spherical scatterer (see Fig.~\ref{fig:m2_0}(a)).
Referring to Fig.~\ref{fig:m2_0},
this is the case where, 
similar to the focal plane
of the incident beam,
the bulk part of the four-peak structure of the focusing zones
is localized inside of the particles.

For $\vc{R}_d=(0,0,\lambda)$,
the focal plane, $z=\lambda$,
is tangential to the shadow part of the particle surface
(see Fig.~\ref{fig:m2_1}(a)).
From Fig.~\ref{fig:m2_1}(b),
it is seen that, as opposed to the case
with $|\vc{R}_d|=0$,
the four peaks of light intensity
now develop in the immediate vicinity of the scatterer surface.

\begin{figure*}[!tbh]
%\centering
%\resizebox{100mm}{!}{\includegraphics*{nanojet.eps}}
\includegraphics[width=100mm]{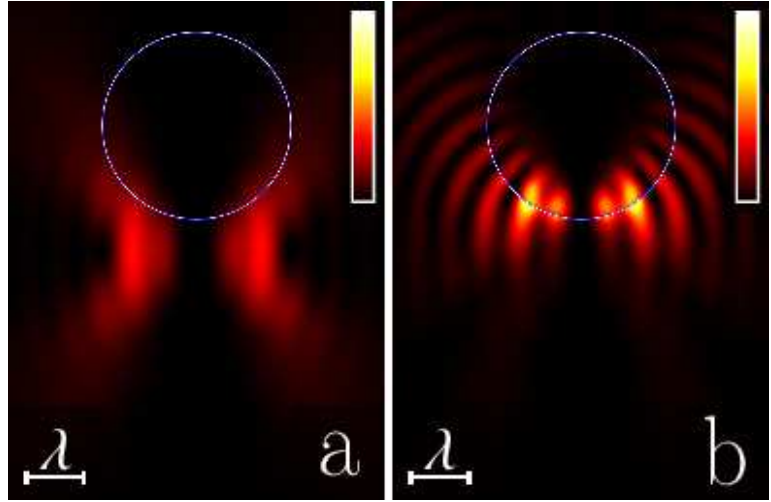}
\caption{%
Near-field intensity distribution in the $x-z$ plane
of  (a)~the incident wave beam
and (b)~the total wave field
for the LG beam with
$m_{\ind{LG}}=3$, $f=0.25$
and $\vc{R}_d=(0,0,\lambda)$. 
}
\label{fig:m3}
\end{figure*}

What all the wavefields depicted in Figs.~\ref{fig:m1}(b)-~\ref{fig:m2_1}(b)
have in common is that,
by contrast to the incident optical vortex beams
with $m_{\ind{LG}}=1$ and $m_{\ind{LG}}=2$,
the light intensity 
on the incident beam axis (the $z$ axis)
clearly differs from zero
(see the neighborhood of the point $(0,0,\lambda)$).
In other words,
it is turned out that,
in the near-field region,
the optical vortex
with $0<|m_{\ind{LG}}|\le 2$ 
has been destroyed
by Mie scattering.
From Fig.~\ref{fig:m3} it is clear that this is no longer the case
at $m_{\ind{LG}}=3$.
This result will be explained in the subsequent section.

%%%%%%%%%%%%%%%%%%%%%%%
\subsection{Optical vortices in near-field region}
\label{subsec:vortices}
%%%%%%%%%%%%%%%%%%%%%%

In this section we consider 
optical vortices and their near-field structure.
The optical vortices are known to represent
phase singularities of complex-valued scalar waves
which are zeros of the wavefield
$\psi=|\psi|\exp(i\chi)$
where its phase $\chi$ is
undefined. 
A phase singularity is 
characterized by the topological vortex charge
$m_{V}$ defined as the closed loop contour
integral of the wave phase $\chi$ modulo $2\pi$
\begin{align}
  \label{eq:vortex-ind}
  m_{V}=\frac{1}{2\pi}\,\oint_L \dd\chi,
\end{align}
where $L$ is the closed path around the singularity.

Optical vortices associated with the individual components of electric field
will be of our primary concern.
More specifically, we shall examine the optical vortex structure of
the components $E_z$ and $E_x$ in the planes $z=z_0$ 
parallel to the $x-y$ plane. 
Since, in such planes, circles naturally play the role of closed loops,
the starting point of our analysis is the electric field vector
expressed as a function of the azimuthal
angle $\phi$ in the following form:
\begin{align}
&
  \label{eq:E-tot-phi}
  \vc{E}=\sum_{\mu=-2}^{2} \vc{E}_{\mu}\exp[i(m_{\ind{LG}}+\mu)\phi]
\\
&
\label{eq:E_mu}
\vc{E}_{\pm 2}\parallel \uvc{x}\mp i \uvc{y},\quad
\vc{E}_{\pm 1}\parallel \uvc{z},\quad
\vc{E}_{0}\perp \uvc{z}.
\end{align}
This formula gives the $\phi$ dependence of 
electric field expansion~\eqref{eq:E_alpha}
in which the coefficients are of the form given by 
Eq.~\eqref{eq:alpha-beta-inc-rules}.
An immediate consequence of Eq.~\eqref{eq:E-tot-phi} 
is that $\vc{E}_{\mu}$ can be different from zero on 
the $z$ axis, $\vc{E}_{\mu}(0,0,z)\ne 0$, 
only if $m_{\ind{LG}}+\mu=0$.

From Eq.~\eqref{eq:E_mu}, at $|m_{\ind{LG}}|=1$, the electric field non-vanishing at the beam
axis is linearly polarized along the $z$ axis, whereas it is circular
polarized at $|m_{\ind{LG}}|=2$.  
The intensity distributions shown in Figs~\ref{fig:m0}--~\ref{fig:m2_1}
clearly indicate that, as opposed to the case with $m_{\ind{LG}}=3$ 
(see Fig.~\ref{fig:m3}),
the $z$ axis is not entirely in the dark region
provided that $0 \le m_{\ind{LG}}<3$.

At $|m_{\ind{LG}}|\ge 3$ and $|\mu|\le 2$, a sum $m_{\ind{LG}}+\mu$ 
cannot be equal to zero and the beam axis is
always a nodal line for the components of electric field.  For 
two-dimensional (2D)
electric field distributions in planes normal to the $z$ axis, it
implies that there is an optical vortex located at the origin.

\begin{figure*}[!tbh]
%\centering
%\resizebox{100mm}{!}{\includegraphics*{nanojet.eps}}
\includegraphics[width=150mm]{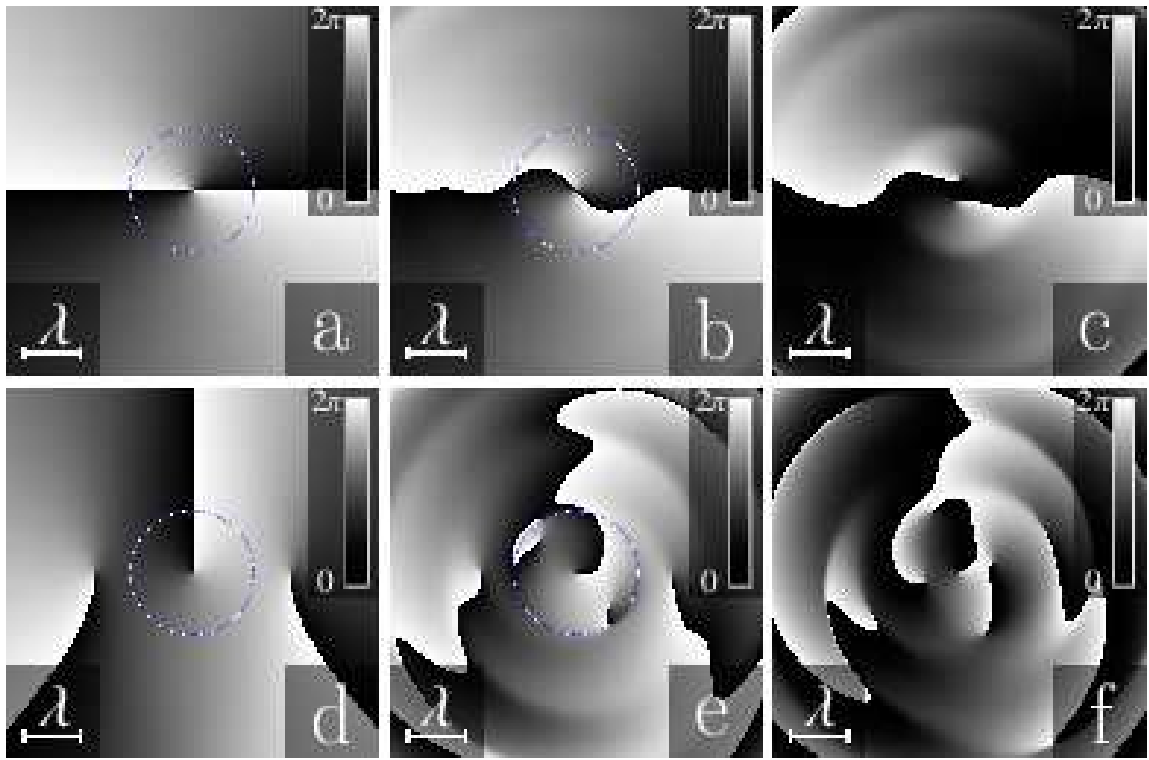}
\caption{%
Near-field phase maps of the electric field components
$E_x$ (a,b,c) and $E_z$ (d,e,f)
in the planes 
$z=0$ (a,b,d,e)
and $z=R_p$ (c,f)
for the LG beam with
$m_{\ind{LG}}=2$ and $f=0.1$.
(a) [(d)]~Phase map of the electric field component
$E_x^{(\ind{LG})}$ [$E_z^{(\ind{LG})}$] 
of the incident wave beam
in the $x-y$ plane ($z=0$).
(b,c) [(e,f)]~Phase maps for the electric field component 
$E_x$ [$E_z$] of 
the total light wavefield in the planes $z=0$ and $z=R_p$, respectively. 
}
\label{fig:phase_m2}
\end{figure*}

Now we turn back to the optical vortex structure
for the components $E_z$ and $E_x$.
The $\phi$ dependence of $E_z$
can be written in the following form:
\begin{align}
&
  \label{eq:E_z}
  \exp[-i m_{\ind{LG}}\,\phi] E_z=
  \exp[-i m_{\ind{LG}}\,\phi+i\chi_z] |E_z|=
E_{+1}^{(z)}\exp[i\phi]+E_{-1}^{(z)}\exp[-i\phi]=
\notag
\\
&
\exp[i\psi_{+}^{(z)}]
\left\{
|E_{+1}^{(z)}|\exp[i(\phi+\psi_{-}^{(z)})]+|E_{-1}^{(z)}|\exp[-i(\phi+\psi_{-}^{(z)})]
\right\},
\end{align}
where 
$E_{\pm 1}^{(z)}=\sca{\vc{E}_{\pm 1}}{\uvc{z}}$,
$2\psi_{\pm}^{(z)}=\arg (E_{+1}^{(z)})\pm\arg (E_{-1}^{(z)})$
and $\chi_z$ is the phase of $E_z$.

In the complex plane formula~\eqref{eq:E_z} describes an ellipse parametrized by
the azimuthal angle $\phi$. It is centered at the origin with the
major (minor) semiaxis of the length $E_{+}^{(z)}(R)$
($|E_{-}^{(z)}(R)| $), where 
$E_{\pm}^{(z)}(R)=|E_{+1}^{(z)}(R)|\pm |E_{-1}^{(z)}(R)|$
$R$ is the radius of circle $C_R$ in the plane of observation,
$z=z_0$.
Then the closed loop contour integral
of the wave phase $\chi_z$ is
\begin{subequations}
 \label{eq:ind-z}
\begin{align}
&
  \label{eq:m-z}
  m_{z}=\frac{1}{2\pi}\,\oint_{C_R} \dd\chi_z=m_{\ind{LG}}+\mu_z(R),
\\
&
  \label{eq:mu-z}
\mu_z(R)=\sign (E_{-}^{(z)}(R))=\sign (|E_{+1}^{(z)}(R)|-|E_{-1}^{(z)}(R)|).
\end{align}
\end{subequations}

\begin{figure*}[!htb]
%\centering
%\resizebox{100mm}{!}{\includegraphics*{nanojet.eps}}
\includegraphics[width=150mm]{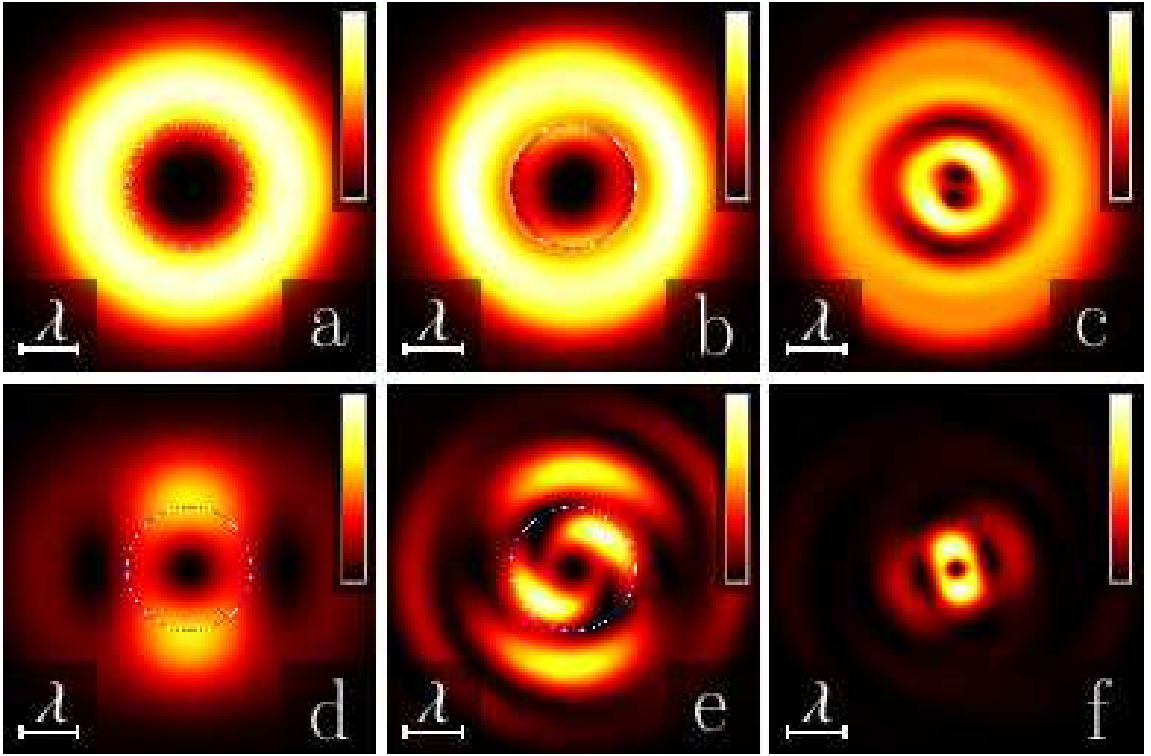}
\caption{%
Near-field intensity maps of the electric field components
$|E_x|^2$ (a,b,c) and $|E_z|^2$ (d,e,f)
in the planes 
$z=0$ (a,b,d,e)
and $z=R_p$ (c,f)
for the LG beam with
$m_{\ind{LG}}=2$ and $f=0.1$.
(a) [(d)]~Intensity distribution for the $x$ [$z$] component,
$|E_x^{(\ind{LG})}|^2$ [$|E_z^{(\ind{LG})}|^2$], 
of the incident wave beam
in the $x-y$ plane ($z=0$).
(b,c) [(e,f)]~Intensity distributions for the $x$ [$z$] component
of electric field of 
the total light wavefield in the planes $z=0$ and $z=R_p$, respectively. 
}
\label{fig:phase_m2_I}
\end{figure*}

From Eq.~\eqref{eq:ind-z}
the net topological charge of vortices encircled by $C_R$ can be
either $m_{\ind{LG}}+1$ or $m_{\ind{LG}}-1$. 
At $|E_{+1}^{(z)}(R)|=|E_{-1}^{(z)}(R)|$, 
$\mu_z(R)$ is undefined. This is the special case
when $|E_z|=0$ at $\cos(\phi+\psi_{-}^{(z)})=0$
and  the circle contains a pair of 
symmetrically located vortices.
Each of these vortices carries
the charge of the magnitude equal to unity. Generally, the vortices
are of the same sign which is determined by the change of $\mu_z(R)$ as the
radius $R$ passes the critical value.  When $\mu_z(R)$ 
changes from $+1$ ($-1$) to
$-1$ ($+1$) two vortices of the charge $-1$ ($+1$) 
intersect the boundary and
move into the interior part of the circle.

The near field phase maps for $\chi_z$
are presented in Figs.~\ref{fig:phase_m2}(d)--(f).
Figure~\ref{fig:phase_m2}(d)
shows the 2D map for the incident optical vortex 
LG beam with $m_{\ind{LG}}=2$ in the focal plane $z=0$.
The corresponding intensity map is depicted in Fig.~\ref{fig:phase_m2_I}(d).
It is seen that there is a vortex
of the charge $m_{\ind{LG}}-1=1$ at the center,
so that, at sufficiently small $R$, $m_z=1$ and $\mu_z=-1$.
In addition, there is 
a pair of the symmetrically arranged
vortices of the charge $+1$ outside the particle.
So, when the radius $R$ is large enough for 
the circle to enclose the three vortices,
the total charge is $m_z=m_{\ind{LG}}+1=3$ and $\mu_z=1$.

For the total wavefield at $z=0$,
the phase and intensity maps are given in
Fig.~\ref{fig:phase_m2}(e) and Fig.~\ref{fig:phase_m2_I}(e),
respectively.
It can be seen that
the vortex pattern is complicated by
interference between the incident and the scattered waves.
Referring to Fig.~\ref{fig:phase_m2}(e),
there are two additional pairs of vortices
whose charges are opposite in sign.
The positively charged vortices 
(the charge is $+1$)
are located inside the particle,
whereas the negatively charged ones 
(the charge is $-1$)
are formed at the surface of the particle.
Similar structure is discernible from
Figs.~\ref{fig:phase_m2}(f) and~\ref{fig:phase_m2_I}(f)
representing the results for the plane 
tangent to the particle surface $z=R_p$.

The case of the $x$ component of the electric field,
$E_x$,
can be analyzed along similar lines.
From Eq.~\eqref{eq:E-tot-phi}, we deduce the $\phi$ dependence of
$E_x$ in the form:
\begin{align}
&
  \label{eq:E_x}
\exp[-i m_{\ind{LG}}\,\phi+i\chi_x] |E_x|-E_{0}^{(x)}=
 E_{+2}^{(x)}\exp[2 i \phi]+E_{-2}^{(x)}\exp[-2 i\phi]=
\notag
\\
&
\exp[i\psi_{+}^{(x)}]
\left\{
|E_{+2}^{(x)}|\exp[i(2 \phi+\psi_{-}^{(x)})]+|E_{-2}^{(x)}|\exp[-i(2 \phi+\psi_{-}^{(x)})]
\right\},
\end{align}
where 
$E_{\pm 2,\,0}^{(x)}=\sca{\vc{E}_{\pm 2,\,0}}{\uvc{x}}$,
$2\psi_{\pm}^{(x)}=\arg (E_{+2}^{(x)})\pm\arg (E_{-2}^{(x)})$
and $\chi_x$ is the phase of $E_x$.
The center of the ellipse described by Eq.~\eqref{eq:E_x}
is generally displaced from the origin
and is determined by $E_0^{(x)}$.
The length of its major (minor) semiaxis is 
$E_{+}^{(x)}(R)$
($|E_{-}^{(x)}(R)| $), where 
$E_{\pm}^{(x)}(R)=|E_{+2}^{(x)}(R)|\pm |E_{-2}^{(x)}(R)|$.

The closed loop contour integral
of the wave phase $\chi_x$ is
\begin{align}
  \label{eq:m-x}
  m_{x}=\frac{1}{2\pi}\,\oint_{C_R} \dd\chi_x=m_{\ind{LG}}+\mu_x(R),
\quad
\mu_x(R)\in\{-2,0,2\}.
\end{align}

When the origin is enclosed by the
ellipse~\eqref{eq:E_x},
similar to Eq.~\eqref{eq:mu-z},
we have the relation
\begin{align}
  \label{eq:mu-x}
\mu_x(R)= 2 \sign (E_{-}^{(x)}(R))=2 \sign (|E_{+2}^{(x)}(R)|-|E_{-2}^{(x)}(R)|).
\end{align}
In the opposite case where the origin is outside the area
encircled by the ellipse, $\mu_x(R)$ is zero.
The latter is the case for the phase maps
representing the 2D distributions of $\chi_x$
in the $x-y$ plane (see Figs.~\ref{fig:phase_m2}(a)--(b)).
As is evident from Figs.~\ref{fig:phase_m2}(a)--(b)
(see also the intensity maps in Figs.~\ref{fig:phase_m2_I}(a)--(b)),
in these distributions, the only vortex
is positioned at the center and possesses the charge 
$m_x=m_{\ind{LG}}=2$.

For the origin located on the ellipse, we generally have
the circle $C_R$ containing
a pair of symmetrically arranged and identically charged
vortices each with the charge magnitude equal to unity.
Note that, by contrast to the case of $E_z$ 
where the origin is placed at the center of the ellipse~\eqref{eq:E_z},
 intersection of $C_R$ and the vortices generally occurs
at non-vanishing $E_{-}^{(x)}$,
$E_{-}^{(x)}\ne 0$,
when
the ellipse~\eqref{eq:E_x} is not degenerated into the interval.

In Fig.~\ref{fig:phase_m2}(c),
we show what happen to 
the above discussed central vortex
in the tangent plane of the particle surface, $z=R_p$.
From the phase and intensity maps
(see Figs.~\ref{fig:phase_m2}(c) and~\ref{fig:phase_m2_I}(c)),
the central vortex has been destroyed and
is replaced by a pair of positively charged and symmetrically arranged
vortices. In this vortex pattern,
$\mu_x(R)=-2$ at small $R$, whereas
$\mu_x(R)$ becomes zero when the vortices are encircled by $C_R$.
For the ellipse~\eqref{eq:E_x}, it means that 
the origin, which is initially encompassed by the ellipse with
$E_{-}^{(x)}<0$,
intersects the ellipse and 
moves outward the area bounded by the ellipse
as $R$ increases.

%%%%%%%%%%%%%%%%%%%%%%%
\section{Conclusions}
\label{sec:conclusions}
%%%%%%%%%%%%%%%%%%%%%%

In this paper, 
we have used
a modified $T$--matrix approach~\cite{Kis:pre:2002}
to study the light scattering problem
for optically isotropic spherical scatterers
illuminated with LG beams 
that represent optical vortex laser 
beams.
In our approach, such beams are described
in terms of the far-field angular distribution~\eqref{eq:E_out} 
using the remodelling procedure
in which the far-field matching method is
combined with  the results for nonparaxial propagation of
LG beams (see Eq.~\eqref{eq:E_ff_LG}).

The analytical results are employed to
perform numerical analysis of
the optical field in the near-field region.
In order to examine the effects of 
incident beam spatial structure on the light wavefield
near the scatterer,
we have computed a number of 
the 2D near-field intensity and phase
distributions for purely azimuthal LG beams.
In this case, a LG beam
possesses the vanishing radial mode number 
and carries the optical vortex
with the topological charge characterized by the azimuthal 
number $m_{\ind{LG}}$.

The 2D near-field intensity distributions
computed for the plane-wave limiting case
in which the incident wave is 
a Gaussian beam ($m_{\ind{LG}}=0$)
with small focusing parameter $f$
($2\pi f=\lambda/w_0< 1$)
reveal  the well-known structure of 
photonic nanojets (see Fig.~\ref{fig:m0}).
Figures~\ref{fig:m1}--~\ref{fig:m3} represent
the results for the LG beams with $1 \le m_{\ind{LG}}\le 3$
and illustrate the following effects:
\renewcommand{\theenumi}{\alph{enumi}}
\renewcommand{\labelenumi}{(\theenumi)}
\begin{enumerate}
\item 
a jetlike photonic 
flux emerging from the particle shadow surface
can be formed
even if the bulk part of the scatterer
is in the low intensity region
(see Fig.~\ref{fig:m1}(b));

\item 
the morphology of photonic jets
formed at $m_{\ind{LG}}\ne 0$
significantly differs from
the well-known shape of nanojet
at  $m_{\ind{LG}}=0$
(see Figs.~\ref{fig:m2_0}--~\ref{fig:m3});

\item 
by contrast to the case with $m_{\ind{LG}}=3$, 
at $|m_{\ind{LG}}|<3$,
the intensity of scattered wavefield
does not vanish on the beam axis 
so that, in the near-feld region, 
lght scattering has a destructive effect
on the optical vortex
(see Figs.~\ref{fig:m1}--~\ref{fig:m2_1}).
\end{enumerate}

Our analysis of optical vortices associated with 
the electric field components
is based on general formula~\eqref{eq:E-tot-phi} 
giving the electric field vector
expressed as a function of the azimuthal angle $\phi$.
Using analytical expressions~\eqref{eq:E_z} and~\eqref{eq:E_x},
we have described the geometry of optical vortices for the components
$E_z$ and $E_x$ in the planes $z=z_0$ normal to the beam axis
(the $z$ axis).

It was found that, except for the central vortex,
the topological charge
of off-center vortices generally equals unity in magnitude.
They are organized into pairs of
symmetrically arranged and equally charged
vortices. These pairs lie on concentric circles
and their vortex charge alternate in sign with the circle radius.

The phase maps of $E_x$  shown in Figs.~\ref{fig:phase_m2}(a)-(c)
(the corresponding square amplitude distributions are presented in 
Figs.~\ref{fig:phase_m2_I}(a)-(c)) are computed for the LG beam
with $m_{\ind{LG}}=2$.
It turned out that
the central vortex of the charge equal to the azimuthal number
$m_{\ind{LG}}=2$ is the only vortex in the $x-y$ plane ($z=0$)
for both the incident beam 
(see Fig.~\ref{fig:phase_m2}(a))
and the total wavefield
(see Fig.~\ref{fig:phase_m2}(b)).
At $z=R_d$, this vortex 
breaks down into a pair of vortices 
each of the unity charge $m_{V}=1$
(see Fig.~\ref{fig:phase_m2}(c)).

By contrast to the case of $E_x$, 
Eq.~\eqref{eq:E_z}
implies that
the $z$ axis is a nodal line 
for the $z$ component of the electric field $E_z$
and
the central vortex is structurally stable
at $m_{\ind{LG}}=2$
(see Figs.~\ref{fig:phase_m2}(d)-(f)).
A comparison between
the phase maps for the incident beam 
(see Fig.~\ref{fig:phase_m2}(d))
and for the total light field 
(see Fig.~\ref{fig:phase_m2}(e))
shows that, in the $x-y$ plane,
interference
between the incident and the scattered waves
produces two additional pairs of vortices.

In conclusion, 
an important consequence of formula~\eqref{eq:E-tot-phi} is
that, at sufficiently large azimuthal numbers, 
$|m_{\ind{LG}}|\ge 3$,
light scattering of LG beams takes place 
without destroying
the optical vortex located on the beam axis. 

% \appendix

%\bibliographystyle{apsrev}
%\bibliographystyle{apsrev4-1}
%\bibliographystyle{lc}
%\bibliography{optics,polymer,scatter,lc,quant,hk,flc,qft,math}

%merlin.mbs apsrev4-1.bst 2010-07-25 4.21a (PWD, AO, DPC) hacked
%Control: key (0)
%Control: author (8) initials jnrlst
%Control: editor formatted (1) identically to author
%Control: production of article title (-1) disabled
%Control: page (0) single
%Control: year (1) truncated
%Control: production of eprint (0) enabled
%

\end{document}